\documentclass[12pt,noshowpacs,nofootinbib,notitlepage,amsmath,superscriptaddress]{revtex4-1}
\usepackage{setspace}
\usepackage{amsfonts}
\usepackage{appendix}
\usepackage{mathtools}
\usepackage{ulem}
\usepackage{ifthen}
\newboolean{PrintVersion}
\setboolean{PrintVersion}{false}
\usepackage{amsmath,amssymb,amstext} 
\usepackage[pdftex]{graphicx} 
\usepackage[caption=false]{subfig}
\usepackage{wrapfig}
\usepackage{cleveref} 
\usepackage{enumitem}
\usepackage{float}
\usepackage{mathtools}
\usepackage{xcolor}
\usepackage[export]{adjustbox}
\usepackage{float}
\usepackage{booktabs}

\ifthenelse{\boolean{PrintVersion}}{   
\hypersetup{	
    citecolor=black,%
    filecolor=black,%
    linkcolor=black,%
    urlcolor=black}
}{} 

\let\origdoublepage\cleardoublepage
\newcommand{\clearemptydoublepage}{%
  \clearpage{\pagestyle{empty}\origdoublepage}}
\let\cleardoublepage\clearemptydoublepage

\begin{document}

\title{Quark stars with a unified interacting equation of state in regularized 4D Einstein-Gauss-Bonnet gravity}

\author{Michael Gammon}
\email{mgammon@uwaterloo.ca}
\affiliation{Department of Physics and Astronomy, University of Waterloo, Waterloo, Ontario, Canada, N2L 3G1}

\author{Sarah Rourke}
\email{sarourke@uwaterloo.ca}
\affiliation{Department of Physics and Astronomy, University of Waterloo, Waterloo, Ontario, Canada, N2L 3G1}
\affiliation{Department of Physics, McGill University, Montreal, Quebec, Canada}

 \author{Robert B. Mann}
\email{rbmann@uwaterloo.ca}
\affiliation{Department of Physics and Astronomy, University of Waterloo, Waterloo, Ontario, Canada, N2L 3G1}

\begin{abstract}
Since the derivation of a well-defined $D\rightarrow 4$ limit for 4D Einstein Gauss-Bonnet (4DEGB) gravity \cite{hennigar_2020_on} coupled to a scalar field, there has been interest in testing it as an alternative to Einstein's general theory of relativity. Using the Tolman-Oppenheimer-Volkoff (TOV) equations modified for 4DEGB gravity, we model the stellar structure of quark stars   using a novel interacting quark matter equation of state \cite{zhang_2021_unified}. We find that increasing the Gauss-Bonnet coupling constant $\alpha$ or the interaction parameter $\lambda$ both tend to increase the mass-radius profiles of quark stars described by this theory, allowing a given central pressure to support larger quark stars  in general. These results logically extend to cases where $\lambda < 0$, in which increasing the magnitude of the interaction effects instead diminishes masses and radii. We also analytically identify a critical central pressure in both regimes, below which no quark star solutions exist due to the pressure function having no roots. Most interestingly, we find that quark stars can exist below the general relativistic Buchdahl bound and Schwarzschild radius $R=2M$, due to the lack of a mass gap between black holes and compact stars in the 4DEGB theory.  Even for small $\alpha$ well within current observational constraints, we find that   quark star solutions in this theory can describe Extreme Compact Objects (ECOs), objects whose radii are smaller than what is allowed by   general relativity.
\end{abstract}

\maketitle

\newpage

\section{Introduction}

Modified theories of gravity  continue to attract attention despite the empirical success of general relativity (GR). These theories are motivated by a variety of problems, including addressing issues in in modern cosmology, quantizing gravity, eliminating singularities, and, perhaps most importantly, finding viable phenomenological competitors against which GR can be tested in the most stringent manner possible.

Higher curvature theories (or HCTs) are amongst the most popular 
modifications. An HCT modifies the assumed linear relationship in GR between the curvature and the stress-energy, replacing the former with  an arbitrary sum of powers of the curvature tensor  (appropriately contracted to two indices). Such modifications could conceivably  improve the empirical success of GR, while also making new testable predictions.

Many quantum gravity proposals \cite{ahmed2017} and cosmological puzzles (such as dark energy, dark matter, and early-time inflation \cite{bueno2016}) suggest that HCTs could play an important role in physics. Lovelock theories \cite{lovelock1971} have long been at the forefront of this search, since they posses the distinctive feature of having 2nd order differential equations of motion. The physical significance of such theories has been unclear, however, since their higher order terms yield non-trivial contributions to the equations of motion only in more than four  spacetime dimensions  ($D>4$). 

Recently this restriction was circumvented in the quadratic case, or what is better known as  ``Einstein-Gauss-Bonnet" gravity.  The  Gauss-Bonnet (GB) contribution to the gravitational action is
\begin{equation}\label{eq:dgbaction}
S_{D}^{G B} = \alpha \int d^{D} x \sqrt{-g} \left[R^{\mu\nu\rho\tau}R_{\mu\nu\rho\tau} - 4 R^{\mu\nu}R_{\mu\nu} + R^2 \right]
\equiv \alpha \int d^{D} x \sqrt{-g} \mathcal{G}
\end{equation}
(where  $R_{\mu\nu\rho\tau}$ is the Riemann curvature tensor), which becomes the integral of a total derivative in $D=4$, and thus cannot  contribute to a system's gravitational dynamics in less than five dimensions. For this reason it is often referred to as a ``topological term" having no relevance to physical problems.  Indeed, the  Lovelock theorem \cite{lovelock1971} ensures that a $D=4$ dimensional metric theory of gravity must incorporate additional fields in order to have second order equations of motion and diffeomorphism invariance.  

Recently it was noted  \cite{glavan2020} that several exact solutions to $D$-dimensional Einstein-Gauss Bonnet gravity have a sensible limit under the rescaling 
\begin{equation}\label{eq:alpharescale}
	\lim_{D \to 4} (D-4) \alpha \rightarrow \alpha,
\end{equation}
of the Gauss-Bonnet coupling constant.  Using this approach a variety of 4-dimensional metrics can be obtained, including cosmological  \cite{glavan2020,li2020,kobayashi2020}, spherical black hole \cite{glavan2020,kumar2022,fernandes2020,kumar2020,kumar2022_2}, collapsing \cite{malafarina2020}, star-like  \cite{doneva2021,charmousis2022}, and radiating  \cite{ghosh2020} metrics, each carrying imprints of the quadratic curvature effects of  their $D > 4$ counterparts.  However a number of objections to this approach were subsequently raised \cite{gurses2020,ai2020,shu2020}, based on the fact that 
the existence of a limiting solution does not imply the existence of a well-defined 4D theory whose field equations have that solution. This shortcoming was quickly addressed  when it was shown that the $D\to 4$ limit in \eqref{eq:alpharescale} can be taken in the gravitational action \cite{hennigar_2020_on,Fernandes:2020nbq}, generalizing an earlier  procedure employed in obtaining the $D\to 2$ limit of GR \cite{Mann:1992ar}. One can also   compactify  $D$-dimensional Gauss-Bonnet gravity on a $(D-4)$-dimensional maximally symmetric space and then use \eqref{eq:alpharescale} to obtain a
$D=4$ HCT \cite{Lu:2020iav}. This approach yields the same result (up to trivial field redefinitions), in addition to terms depending on the curvature of the maximally symmetric $(D-4)$-dimensional space. Taking this to vanish yields
\begin{equation}\label{eq:4DEGBaction}
\begin{aligned}
S_{4}^{GB}
=&\alpha \int d^{4} x \sqrt{-g}\left[  \phi \mathcal{G}+4 G_{\mu \nu} \nabla^\mu \phi \nabla^\nu \phi-4(\nabla \phi)^2 \square \phi+2(\nabla \phi)^4\right]
\end{aligned}
\end{equation}
where we see that an additional scalar field $\phi$ appears. Surprisingly, the spherically symmetric black hole solutions to the field equations match those from the naïve $D \rightarrow 4$ limit of solutions \cite{glavan2020}. The resultant 4D scalar-tensor theory is a particular type of Horndeski theory \cite{horndeski1974}, and solutions to its equations of motion can be obtained without ever referencing a higher dimensional spacetime \cite{hennigar_2020_on}. 

We are interested here in what is called 4D Einstein-Gauss-Bonnet gravity (4DEGB), whose action is given by \eqref{eq:4DEGBaction} plus the Einstein-Hilbert term:

\begin{equation}
\begin{aligned}
S= S^{GR} + S_{4}^{GB}
=& \int d^{4} x \sqrt{-g}\left[R + 
\alpha\left\{  \phi \mathcal{G}+4 G_{\mu \nu} \nabla^\mu \phi \nabla^\nu \phi-4(\nabla \phi)^2 \square \phi+2(\nabla \phi)^4 \right\}\right]
\end{aligned}
\label{4DEGB}
\end{equation}
which has been shown to be   an interesting phenomenological competitor to GR \cite{Clifton:2020xhc}. 
Despite much exploration \cite{Fernandes:2022zrq}, it is unclear whether these higher curvature terms play an important role in real gravitational dynamics. One important arena for testing such theories against standard general relativity is via observations of compact astrophysical objects like neutron stars. The correct theory should be able to accurately describe recent gravitational wave observations of astrophysical objects existing in the mass gap between the heaviest compact stars and the lightest black holes. 

 Modern observational astrophysics is rich in findings of compact objects and as such our understanding of highly dense gravitational objects is rapidly advancing. However, there is as of yet no strong consensus on their underlying physics. 
 A number of such objects have been recently observed that are inconsistent with standard GR and a simple neutron star equation of state. It was recently shown \cite{zhang_2021_unified,zhang_2021_stellar} that in standard general relativity, the secondary component of the merger GW190814 could feasibly be a quark star with an interacting equation of state governed by a single parameter $\lambda$.  This parameterization of strong interaction effects was inspired by another recent theoretical study showing that non-strange quark matter could feasibly be the ground state of baryonic matter at sufficient density and temperature \cite{holdom_2018_quark}. Similar analyses with a different equation of state and/or QM phase \cite{miao_2021_bayesian,lopes_2022_onthe,oikonomou_2023_colourflavour} found similarly promising results.
This same object was subsequently shown to be well described as a slowly-rotating neutron star in the 4DEGB theory without resorting to exotic quark matter EOSs, while also demonstrating that the equilibrium sequence of neutron stars asymptotically matches the black hole limit, thus closing the mass gap between NS/black holes of the same radius \cite{charmousis2022}.  More recently, some groups have also been interested in modelling ECOs \cite{zhang_2023_rescaling} as well as unusually light compact stars \cite{horvath_2023_alight,oikonomou_2023_colourflavour} (like that in the gamma ray remnant J1731-347, which is inconsistent with minimum mass calculations of neutron stars generated by iron cores) as quark stars to explain their unusual properties. 


 To further illuminate the range of possibilities, we consider in this paper quark star solutions to the 4DEGB theory. Inspired by the unified interacting quark matter equation of state derived in \cite{zhang_2021_unified}, we present a simple model for interacting quark stars in the regularized 4DEGB theory in which all of the strong interaction effects (including corrections from perturbative QCD, colour superconductivity, and the strange quark mass) are characterized by a single  parameter $ {\lambda}$ in the equation of state. In doing so, we solve the modified TOV equations for a number of combinations of Gauss-Bonnet coupling and QM interaction strengths (constrained by observational limits). Although quark star solutions have previously been considered in the context of 4DEGB \cite{banerjee_2021_strange,banerjee_2021_quark,pretel_2022}, a
different (less general)   equation of state was used, and the upper limit of the coupling $\alpha$ was taken to be much smaller than that allowed by current observational constraints \cite{Fernandes:2022zrq,Clifton:2020xhc}  and the 4DEGB Buchdahl bound \cite{chakraborty2020_limits}. In considering 
a unified, interacting equation of state \cite{zhang_2021_unified} and values of $\alpha$ up to presently allowed bounds, we obtain a number of interesting novel results. 

Our most intriguing result is that quark stars in 4DEGB can be Extreme Compact Objects (ECOs), objects whose radii are smaller than that allowed by the Buchdahl bound in GR. Indeed, there exist quark stars in 4DEGB whose radii are smaller than that of a corresponding black hole of the same mass in GR\footnote{This phenomenon was also shown to be present for neutron stars
\cite{charmousis2022}, though the relationship with the Buchdahl bound was not noted.}. Observations of these latter objects, apart from indicating a new class of astrophysical phenomena \cite{Mann:2021mnc}, would provide strong evidence for 4DEGB as a physical theory. These ECOs respect a generalization of the Buchdahl bound, whose small-radius limit is that of the horizon radius of the corresponding black hole.

We find in general that for a given central pressure, quark stars in 4DEGB have larger mass and radius than their GR counterparts with the same pressure.  This can be attributed to the `less attractive' nature of gravity in 4DEGB: for $\alpha>0$ the gradient of the effective potential yields a weaker force than pure GR. We consequently find a larger maximal mass for a given value of $\lambda$ in 4DEGB than in general relativity.

Surprisingly we were also able to analytically derive a critical central pressure  for certain $\alpha/\lambda$ combinations, below which quark stars cannot form,  and above which the mass and radius are greater than or equal to those of a black hole in the 4DEGB theory. With this, it is clear that for such parameter combinations no stable quark star solutions exist. We  analytically derive an expression for this critical quantity in terms of the couplings $\alpha$ and $\lambda$. Moreover, we find that both increasing the 4DEGB coupling constant $\alpha$ and increasing the interaction parameter $\lambda$ make the mass-radius profiles of the stars larger (and consequently lead to a larger maximum mass point for a given parameter set). The converse is also true for smaller values of these parameters, with the trend continuing as $\lambda<0$.

The outline of our paper is as follows: In section \ref{sec:theory} we introduce the basic theory underlying 4DEGB gravity, as well as the unified, interacting quark matter equation of state that we make use of. Following this, a perfect fluid stress-energy tensor is employed to derive the 4DEGB TOV equations, and current observational constraints on the coupling constant are briefly discussed. Section \ref{sec:results} outlines the results of our numerical calculations, with the first part of the section covering $\lambda>0$ solutions (positive interaction strength), and the second part seeing how these results change when $\lambda<0$. We conclude our results with a brief analysis of the stability of unified, interacting 4DEGB quark stars. Section \ref{sec:summary} summarizes our key findings and suggests topics for future study.

\section{Theory}\label{sec:theory}

\subsection{4D Einstein-Gauss-Bonnet Gravity}

The field equations of 4DEGB are obtained from a straightforward variational principle applied to the action \eqref{4DEGB}. Variation with respect to the scalar $\phi$ yields
\begin{equation}\label{eq:eomscalar}
\begin{aligned}
\mathcal{E}_{\phi}=&-\mathcal{G}+8 G^{\mu \nu} \nabla_{\nu} \nabla_{\mu} \phi+8 R^{\mu \nu} \nabla_{\mu} \phi \nabla_{\nu} \phi-8(\square \phi)^{2}+8(\nabla \phi)^{2} \square \phi+16 \nabla^{a} \phi \nabla^{\nu} \phi \nabla_{\nu} \nabla_{\mu} \phi \\
&\qquad +8 \nabla_{\nu} \nabla_{\mu} \phi \nabla^{\nu} \nabla^{\mu} \phi \\
=& \; 0
\end{aligned}
\end{equation}
and the variation with respect to the metric gives
\begin{equation}\label{eq:eommetric}
\begin{aligned}
\mathcal{E}_{\mu \nu} &=\Lambda g_{\mu \nu}+G_{\mu \nu}+\alpha\left[\phi H_{\mu \nu}-2 R\left[\left(\nabla_{\mu} \phi\right)\left(\nabla_{\nu} \phi\right)+\nabla_{\nu} \nabla_{\mu} \phi\right]+8 R_{(\mu}^{\sigma} \nabla_{\nu)} \nabla_{\sigma} \phi+8 R_{(\mu}^{\sigma}\left(\nabla_{\nu)} \phi\right)\left(\nabla_{\sigma} \phi\right)\right.\\
&-2 G_{\mu \nu}\left[(\nabla \phi)^{2}+2 \square \phi\right]-4\left[\left(\nabla_{\mu} \phi\right)\left(\nabla_{\nu} \phi\right)+\nabla_{\nu} \nabla_{\mu} \phi\right] \square \phi-\left[g_{\mu \nu}(\nabla \phi)^{2}-4\left(\nabla_{\mu} \phi\right)\left(\nabla_{\nu} \phi\right)\right](\nabla \phi)^{2} \\
&+8\left(\nabla_{(\mu} \phi\right)\left(\nabla_{\nu)} \nabla_{\sigma} \phi\right) \nabla^{\sigma} \phi-4 g_{\mu \nu} R^{\sigma \rho}\left[\nabla_{\sigma} \nabla_{\rho} \phi+\left(\nabla_{\sigma} \phi\right)\left(\nabla_{\rho} \phi\right)\right]+2 g_{\mu \nu}(\square \phi)^{2} \\
& -4 g_{\mu \nu}\left(\nabla^{\sigma} \phi\right)\left(\nabla^{\rho} \phi\right)\left(\nabla_{\sigma} \nabla_{\rho} \phi\right)+4\left(\nabla_{\sigma} \nabla_{\nu} \phi\right)\left(\nabla^{\sigma} \nabla_{\mu} \phi\right) \\ 
&\left. -2 g_{\mu \nu}\left(\nabla_{\sigma} \nabla_{\rho} \phi\right)\left(\nabla^{\sigma} \nabla^{\rho} \phi\right)
+4 R_{\mu \nu \sigma \rho}\left[\left(\nabla^{\sigma} \phi\right)\left(\nabla^{\rho} \phi\right)+\nabla^{\rho} \nabla^{\sigma} \phi\right] \right]\\
&=\;T_{\mu \nu}
\end{aligned}
\end{equation}
where 
\begin{equation}\label{eq:gbtensor}
    \begin{aligned}
    H_{\mu \nu}=2\Big[R R_{\mu \nu}-2 R_{\mu \alpha \nu \beta} R^{\alpha \beta}+R_{\mu \alpha \beta \sigma} R_{\nu}^{\alpha \beta \sigma}-2 R_{\mu \alpha} R_{\nu}^{\alpha} -\frac{1}{4} g_{\mu \nu}\mathcal{G}
\Big]
    \end{aligned}
\end{equation}
 is  the Gauss-Bonnet tensor. These field equations satisfy the following relationship
\begin{equation}\label{eq:fieldeqntrace}
g^{\mu \nu}T_{\mu \nu}=g^{\mu \nu} \mathcal{E}_{\mu \nu}+\frac{\alpha}{2} \mathcal{E}_{\phi}=4 \Lambda-R-\frac{\alpha}{2} \mathcal{G}
\end{equation}
which can act as a useful consistency check to see whether prior solutions generated via the Glavin/Lin method are even possible solutions to the theory. 
For example, using \eqref{eq:fieldeqntrace}
it is easy to verify that the rotating metrics generated from a Newman-Janis algorithm \cite{kumar2020,wei2020} are not solutions to the field equations of the scalar-tensor 4DEGB theory.

\subsection{Unified Interacting Quark Matter Equation of State}\label{subsec:qstheory}

Compact stars described in terms of deconfined quark degrees of freedom are often modelled using the simple, non-interacting quark matter equation of state \cite{glendenningbook}
\begin{equation}
p(r) = \frac{1}{3}(\rho (r) - 4 B_{\mathrm{eff}})
\end{equation}
where $p$ and $\rho$ are pressure and mass density (respectively), and $B_{\mathrm{eff}}$ is the effective bag constant from the MIT bag model for quark confinement. The bag constant values associated with conventional non-interacting strange quark matter (SQM) or up-down quark matter (\textit{ud}QM) are not small enough to explain the large compact star masses found in recent binary merger events using GR alone \cite{zhang2020,ren2020,zhou_2018}. \\

Accordingly, Zhang and Mann \cite{zhang_2021_unified} investigated whether a strongly interacting equation of state would allow a quark star to fit the observational constraints, particularly those from recent
analysis by   NICER~\cite{Riley:2019yda,Miller:2019cac}, and  binary merger gravitational-wave events~\cite{LIGOScientific:2017vwq,LIGOScientific:2018hze}, 
in the context of GR.  
This EOS was inspired by a recent theoretical development \cite{holdom_2018_quark} showing that \textit{ud}QM could be the ground state of baryonic matter at sufficiently low density and temperature. In employing this, they also provided a unified framework for all strongly interacting phases of quark matter, condensing the entirety of strong interaction effects into a single parameter $\lambda$. This greatly simplifies the problem as it is no longer necessary to solve the same equations for different phases of quark matter (namely the two-flavour superconducting without/with strange quarks (2SC, 2SC+s), and colour-flavour locking (CFL) phases), different values of the gap parameter from colour superconductivity, different perturbative contributions etc. as all these effects are unified in  the model. 

The basic framework in deriving this involves writing the free energy $\Omega$ of the superconducting quark matter in the following form  \cite{alford_2002_absence,fraga_2001_small,alford_2005_hybrid,weissenborn_2011_quark}:
\begin{equation}\label{eq:freenrg}
\begin{aligned}
\Omega= & -\frac{\xi_4}{4 \pi^2} \mu^4+\frac{\xi_4\left(1-a_4\right)}{4 \pi^2} \mu^4-\frac{\xi_{2 a} \Delta^2-\xi_{2 b} m_s^2}{\pi^2} \mu^2 -\frac{\mu_e^4}{12 \pi^2}+B_{\text {eff }}
\end{aligned}
\end{equation}
where $\mu$ and $\mu_e$ average quark and electron chemical potentials respectively, $\Delta$ is the gap parameter, $m_s$ accounts for corrections from the finite strange quark mass (if applicable), $a_4$ represents the 
perturbative Quantum Chromodynamics
(pQCD) contribution from one-gluon exchange, and the constant coefficients
\begin{equation}
\left(\xi_4, \xi_{2 a}, \xi_{2 b}\right)= \begin{cases}\left(\left(\left(\frac{1}{3}\right)^{\frac{4}{3}}+\left(\frac{2}{3}\right)^{\frac{4}{3}}\right)^{-3}, 1,0\right) & \text { 2SC phase } \\ (3,1,3 / 4) & \text { 2SC+s phase } \\ (3,3,3 / 4) & \text { CFL phase }\end{cases}
\end{equation}
account for the various possible QM phases. Since equation \eqref{eq:freenrg} is the grand potential which doesn't account for temperature effects, these expressions are only valid at low temperature. Using the thermodynamic relations
\begin{equation}
p=-\Omega, n_q=-\frac{\partial \Omega}{\partial \mu}, n_e=-\frac{\partial \Omega}{\partial \mu_e}, \rho=\Omega+n_q \mu+n_e
\end{equation}
along with the reparameterization
\begin{equation}
\lambda=\frac{\xi_{2 a} \Delta^2-\xi_{2 b} m_s^2}{\sqrt{\xi_4 a_4}},
\end{equation}
we obtain
\begin{equation}
n_q=\frac{\xi_4 a_4}{\pi^2} \mu^3+\frac{\lambda \sqrt{\xi_4 a_4}}{\pi^2} 2 \mu, \quad n_e=\frac{\mu_e^3}{3 \pi^2},
\end{equation}
\begin{equation}\label{eq:density}
\rho=\frac{3 \xi_4 a_4}{4 \pi^2} \mu^4+\frac{\mu_e^4}{4 \pi^2}+B_{\text {eff }}+\frac{\lambda \sqrt{\xi_4 a_4}}{\pi^2} \mu^2.
\end{equation}

Combining equation \eqref{eq:density} with \eqref{eq:freenrg}, the pressure can be written as a function of mass density $\rho(r)$, the MIT bag constant, and the unified interaction strength $\lambda$: 
\begin{equation}\label{eq:eos}
p(r)=\frac{1}{3}\left(\rho(r)-4 B_{\text {eff }}\right)+\frac{4 \lambda^2}{9 \pi^2}\left(-1+\operatorname{sgn}(\lambda) \sqrt{1+3 \pi^2 \frac{\left(\rho(r)-B_{\text {eff }}\right)}{\lambda^2}}\right)
\end{equation}
where $\rm{sgn}(\lambda)$ represents the sign of $\lambda$. It is easy to see from here how different combinations of parameters may result in the same unified interaction strength $\lambda$, and thus an equivalent equation of state. The above expression can be further generalized by dividing out the $B_{\rm{eff}}$ dependence, leaving us with fully dimensionless equations. Performing the rescalings
\begin{equation}\label{rs1}
\bar{\rho}=\frac{\rho}{4 B_{\mathrm{eff}}}, \bar{p}=\frac{p}{4 B_{\mathrm{eff}}},
\end{equation}
and
\begin{equation}\label{lameff}
\bar{\lambda}=\frac{\lambda^2}{4 B_{\text {eff }}}=\frac{\left(\xi_{2 a} \Delta^2-\xi_{2 b} m_s^2\right)^2}{4 B_{\text {eff }} \xi_4 a_4},
\end{equation}
allows us to rewrite the  EOS \eqref{eq:eos} in terms of these dimensionless parameters as
\begin{equation}\label{eq:eosunitless}
\bar{p}(\bar{r})=\frac{1}{3}(\bar{\rho}(\bar{r})-1)+\frac{4}{9 \pi^2} \bar{\lambda}\left(-1+\operatorname{sgn}(\lambda) \sqrt{1+\frac{3 \pi^2}{\bar{\lambda}}\left(\bar{\rho}(\bar{r})-\frac{1}{4}\right)}\right)
\end{equation}
where
\begin{equation}\label{rs2}
\bar{m}=m \sqrt{4 B_{\mathrm{eff}}}  \qquad \bar{r}=r \sqrt{4 B_{\mathrm{eff}}} \qquad \bar{\alpha} = \alpha \cdot 4 B_{\text {eff}}
\end{equation}
and in the sequel we shall assume a characteristic value of $B_{\mathrm{eff}}= 60\; \rm{MeV/fm^3}$.    

In the limit $\bar{\lambda} \rightarrow 0$, \eqref{eq:eosunitless} reduces back to the expected non-interacting EOS $\bar{p}(\bar{r}) = \frac{1}{3}(\bar{\rho} (\bar{r}) - 1)$, whereas in the limit of extreme positive interaction strength ($\lambda \to +\infty$), \eqref{eq:eosunitless} approaches the form 
\begin{equation}
\left.\bar{p}\right|_{\bar{\lambda} \rightarrow \infty}=\bar{\rho}-\frac{1}{2}
\end{equation}
or, equivalently, $p(r)=\rho(r)-2 B_{\mathrm{eff}}$. In effect, this means that the strong interaction can reduce the surface mass density of the quark star from $\rho_0 = 4 B_{\mathrm{eff}}$ to $\rho_0 = 2 B_{\mathrm{eff}}$, and increase the speed of sound in QM from $\frac{1}{3} c$ to $c$ maximally. On the other hand, when $\lambda$ is negative no well-defined large magnitude limit exists.

We note once more that the  EOS \eqref{eq:eos} 
is applicable only at low temperatures, since it was obtained from the grand potential \eqref{eq:freenrg}, which does  not account for  temperature effects.  This is in contrast to an EOS obtained from  non-perturbative propagators arising from fits to lattice QCD \cite{Canfora:2016xnc}, where results in the  low temperature,  intermediate density regime are not reliable.  In general astrophysical settings,  we are interested in old quark stars, whose temperature is well below the expected value of other mass scales  due to fast cooling.

\subsection{4DEGB TOV Equations}

The standard Tolman-Oppenheimer-Volkoff (TOV) equations for stellar structure are well-known in GR. To model the structure of a quark star in 4DEGB gravity, these relations need to be re-derived. In the following we do this, starting with a static, spherically symmetric metric ansatz in natural units ($G = c = 1$):

\begin{equation}
d s^2=-e^{2 \Phi (r)} c^2 d t^2+ e^{2 \Lambda (r)} d r^2+r^2 d \Omega^2.
\end{equation}

As usual \cite{hennigar_2020_on,gammon_2022}, 
so long as $e^{2\Phi} = e^{2 \Lambda}$ outside the star, the combination $\mathcal{E}_0^0-\mathcal{E}_1^1$ of the field equations can be used to derive the following equation for the scalar field:
\begin{equation}\label{phivac}
    \left(\phi^{\prime 2}+\phi^{\prime \prime}\right)\left(1-\left(r \phi^{\prime}-1\right)^2 e^{-2 \Lambda}\right)=0.
\end{equation}
which, apart from the irrelevant $\phi=\ln \left(\frac{r-r_0}{l}\right)$ (with $r_0$ and $l$ being constants of integration), 
has the solution
\begin{equation}
\phi_{ \pm}=\int \frac{1 \pm e^{\Lambda}}{r} d r
\end{equation}
where  $\phi_-$
falls off as as $\frac{1}{r}$ in asymptotically flat spacetimes.  Choosing 
$\phi = \phi_-$ ensures  that \eqref{eq:eomscalar} is automatically satisfied.

Modelling the quark matter by a perfect fluid matter source, the  stress-energy tensor is
\begin{equation}
T_{\mu \nu}=(\rho+p) u_\mu u_\nu+p g_{\mu \nu},
\end{equation}
from which we obtain the equations
\begin{align}
& \frac{2}{r} \frac{d \Lambda}{d r}=e^{2 \Lambda}\left[\frac{8 \pi G}{c^4} \rho-\frac{1-e^{-2 \Lambda}}{r^2}\left(1-\frac{\alpha\left(1-e^{-2 \Lambda}\right)}{r^2}\right)\right]\left[1+\frac{2 \alpha\left(1-e^{-2 \Lambda}\right)}{r^2}\right]^{-1}, \\
& \frac{2}{r} \frac{d \Phi}{d r}=e^{2 \Lambda}\left[\frac{8 \pi G}{c^4} P+\frac{1-e^{-2 \Lambda}}{r^2}\left(1-\frac{\alpha\left(1-e^{-2 \Lambda}\right)}{r^2}\right)\right]\left[1+\frac{2 \alpha\left(1-e^{-2 \Lambda}\right)}{r^2}\right]^{-1}, \\
& \frac{d p}{d r}=-(\rho+p) \frac{d \Phi}{d r},
\end{align}
in 4DEGB, 
matching those found previously \cite{banerjee_2021_quark}. 

Asymptotic flatness as a condition means that $\Phi(\infty)=\Lambda(\infty)=0$, and regularity at the center of the star implies $\Lambda(0)=0$. If, in the usual way, we define the gravitational mass $m$ within a sphere of radius $r$ through the relation \cite{hennigar_2020_on}
\begin{equation}
\label{Lamdef}
    e^{-2 \Lambda}=1+\frac{r^2}{2 \alpha}\left(1-\sqrt{1+\frac{8 \alpha m(r)}{r^3}}\right)
\end{equation}
 we arrive at the 4DEGB modified TOV equations, namely
\begin{align}
& \frac{d p}{d r}=\frac{(p+\rho)\left[r^3(\Gamma+8 \pi \alpha p-1)-2 \alpha m\right]}{r^2 \Gamma\left[r^2(\Gamma-1)-2 \alpha\right]}\\
&\frac{d m}{d r}=4 \pi r^2 \rho
\label{dmdp}
\end{align}
along with the EOS \eqref{eq:eosunitless}, 
where $\Gamma=\sqrt{1+\frac{8 \alpha m}{r^3}}$. Note that this expression differs from the equivalent equation in \cite{banerjee_2021_quark,banerjee_2021_strange} due to our inclusion of the 4DEGB coupling in the definition of  gravitational mass, however we do find agreement with \cite{pretel_2022}. 

The vacuum solution is given by $m(r) = M$, where $M$ is constant, implying that $\Phi = -\Lambda$. Writing $ e^{-2 \Lambda}=1+2\varphi(r)$, we can   compute the gravitational force in 4DEGB due to a spherical body
\begin{equation}\label{force}
    \vec{F} = -\frac{d\varphi}{dr}\hat{r} = -\frac{r}{2\alpha}
\left(1- \frac{r^3+2\alpha M}{r^3+8\alpha M}\sqrt{1+\frac{8 \alpha M}{r^3}}\right)  \hat{r}
\end{equation}
which is smaller in magnitude than its Newtonian $\alpha=0$ counterpart
($\vec{F}_N = -\frac{M}{r^2}\hat{r}$) for $\alpha>0$. The force in 
\eqref{force} vanishes at 
$r=(\alpha M)^{1/3}$, but this is always at a smaller value of $r$ than the outer horizon 
$R_h = M -\sqrt{M^2 -\alpha}$
of the corresponding black hole. Hence the gravitational force outside of any spherical body, while weaker than that in GR, is always attractive provided $\alpha>0$.  If $\alpha<0$ then the corresponding gravitational force is more attractive than in GR.

Rescaling the various quantities using
\eqref{rs1}, \eqref{lameff}, and \eqref{rs2} we obtain  the unitless equations 
\begin{align}\label{pdiff}
& \frac{d \bar{p}}{d \bar{r}}=\frac{(\bar{p}+\bar{\rho})\left[\bar{r}^3(\Gamma+8 \pi \bar{\alpha} \bar{p}-1)-2 \bar{\alpha} \bar{m}\right]}{\bar{r}^2 \Gamma\left[\bar{r}^2(\Gamma-1)-2 \bar{\alpha}\right]}\\
&\frac{d \bar{m}}{d \bar{r}}=4 \pi \bar{r}^2 \bar{\rho}
\label{mdiff}
\end{align}
which may be solved numerically.
In the limit $\alpha \to 0$, the above equations reduce back to the well-known TOV equations for a static, spherically symmetric gravitating body in GR. \\

To solve \eqref{pdiff} and \eqref{mdiff} numerically we impose the boundary conditions 
\begin{equation}\label{bcs}
m(0)=0, \quad \rho(0)=\rho_{\mathrm{c}},
\end{equation}
where the star's surface radius $R$ is defined via $\bar{p}(\bar{R}) = 0$, namely
 the radius at which the pressure goes to 0 ({\it i.e.} $p(R) = 0$). We similarly define the total mass of the star to be $M = m(R)$. Numerical solutions can thus be obtained by scanning through a range of values of $\rho_c$ and solving for the star's total mass and radius.

Before proceeding to solve the TOV equations, we consider the behaviour of the scalar field $\phi$ in the interior. Inserting the   interior solution \eqref{Lamdef}
 into \eqref{phivac}, we find
\begin{equation}
\begin{aligned}
    \lim _{r \rightarrow 0} \phi^{\prime} \approx & -\sqrt{\frac{m(0)}{2 \alpha r}}-\frac{3 m(0)}{4 \alpha}-\frac{m(0) \sqrt{r}\left(2 \alpha m^{\prime}(0)+5 m(0)^2\right)}{4 \sqrt{2}(\alpha m(0))^{3 / 2}} \\
& -\frac{r\left(8 \alpha\left(3 m^{\prime}(0)-1\right)+35 m(0)^2\right)}{32 \alpha^2}+\mathcal{O}\left(r^{3 / 2}\right) .
\end{aligned}
\end{equation}
Furthermore, provided 
$m(r)$ vanishes at least quadratically in $r$ for small $r$ (which is ensured
from \eqref{dmdp} for the boundary conditions \eqref{bcs})
we find that near the origin
\begin{equation}
\begin{aligned}
    \lim _{r \rightarrow 0} \phi^{\prime} \sim-\sqrt{\frac{r\mathcal{M}(0)} {2\alpha}}+\frac{r}{4 \alpha} \approx 0, \quad \lim _{r \rightarrow 0} \phi \approx K-\frac{\sqrt{2\mathcal{M}(0)}}{3 \sqrt{\alpha}} r^{3/2}+\frac{r^2}{8 \alpha} \approx K,
\end{aligned}
\end{equation}
(where $K$ is a constant and $\mathcal{M}(r)=r^{-2} m(r)$) and thus regularity of the scalar at the origin is ensured.

Finally we note that the  effective bag constant   $B_\mathrm{eff} = 60$ MeV/$\mathrm{fm}^3$  can be converted to units of inverse length squared 
(`gravitational units')
with the factor $G/c^4$, yielding
\begin{equation}
    B_\mathrm{eff} = 7.84 \times 10^{-5} \; \mathrm{km}^{-2}.
\end{equation}

\subsection{Observational Constraints on the 4DEGB Coupling Constant}

A recent study of the observational constraints on the coupling $\alpha$ yielded
\cite{Clifton:2020xhc,Fernandes:2022zrq}
\begin{equation}\label{eq:constraints}
-10^{-30} \textrm{m}^2 < \alpha<10^{10}\; \textrm{m}^2
\end{equation}
where   the lower bound comes from ``early universe cosmology and atomic nuclei"
\cite{charmousis2022}, 
and the upper bound follows from LAGEOS satellite observations. Regarding the lower bound as negligibly close to zero,
the dimensionless version of \eqref{eq:constraints} reads 
\begin{equation}\label{eq:dimlessconstraints}
0<\bar{\alpha} \lesssim 3.2.
\end{equation}

We note that inclusion of preliminary calculations on recent GW data suggest these constraints could potentially tighten   to $0<\alpha \lesssim 10^7\; \textrm{m}^2$, or alternatively $0<\bar{\alpha} \lesssim 0.0032$. This would mean  that deviations from GR due to 4DEGB would only be detectable in extreme environments such as in the very early universe or near the surface of extremely massive objects like black holes.  Even   tighter bounds were assumed in previous studies of quark stars, where only solutions with $\alpha$ below 6 $\mathrm{km}^2$ ($\bar{\alpha} \leq 0.0019$) were considered \cite{banerjee_2021_strange,banerjee_2021_quark,pretel_2022}.
Adopting such a tight bound would make  compact stars near the upper end of the mass gap an ideal candidate for investigation the effects of 4DEGB theory.  

At this point in time such tighter bounds are not warranted.  A proper study of the effects of gravitational radiation in 4DEGB has yet to be carried out.  In view of this we shall assume the bound  \eqref{eq:dimlessconstraints}, which has strong observational support \cite{Clifton:2020xhc,Fernandes:2022zrq}.

\section{Results}\label{sec:results}

In this section we numerically obtain the total quark star mass as a function of  both total star radius and central density, with the former plots being supplemented by the GR/4DEGB black hole horizon radii \cite{hennigar_2020_on,glavan_2020_einsteingaussbonnet} and the GR/4DEGB Buchdahl limits \cite{chakraborty2020_limits}.

Recall that the vacuum solution to 
the field equations 
is given by \eqref{Lamdef} with
$m(r)=M$, a constant, and
$\Phi = -\Lambda$. 
Assuming no star, this 
 solution describes a black hole with two horizons provided $M > \sqrt{\alpha} = M^\mathrm{BH}_\mathrm{min}$. 
The outer horizon radius is
\cite{hennigar_2020_on,glavan_2020_einsteingaussbonnet,gammon_2022}
\begin{equation}\label{eq:bhhorizon}
   R_h = M + \sqrt{M^2-\alpha}\; .
\end{equation}

The Buchdahl bound has been derived in 4DEGB \cite{chakraborty2020_limits} using arguments similar to those in GR, namely that the pressure of the star remains positive (and monotonically decreasing) throughout, and for a given mass the radius of the star must be larger than that of the (outer) horizon of a black hole of the same mass. This yields
\begin{equation}\label{eq:fullbuch4DEGB}
    \sqrt{1-\mu R^2}\left(1+\alpha \mu\right)>\frac{1}{3}\left(1-\alpha \mu \right)
\end{equation}
where 
$$
\mu \equiv \frac{1}{2 \alpha}(\sqrt{1+\frac{8 M \alpha}{R^3}}-1)
$$ 
for a star of radius $R$ and mass $M$. For small $\alpha$, \eqref{eq:fullbuch4DEGB} becomes
\begin{equation}\label{eq:buch4DEGB}
\frac{M}{R} \leq \frac{4}{9}+\frac{16}{27}\left(\frac{\alpha}{R^2}\right)
\end{equation}
and we see that  the GR Buchdahl bound is recovered in the $\alpha \to 0$ limit. This latter relation holds provided spherical symmetry and isotropy are valid, regardless of whether the internal density distribution is constant or a function of the radial coordinate \cite{chakraborty2020_limits}.

Unlike GR, the 4DEGB theory   lacks a mass gap between a compact star and black hole of the same radius.  This feature was first observed in for neutron stars, where the  $M$~vs.~$R$ curves were found to asymptote to the outer black hole horizon for sufficiently large $\alpha$ \cite{charmousis2022}. To find the point at which the Buchdahl bound intersects the black hole horizon, we  substitute  \eqref{eq:bhhorizon} into \eqref{eq:fullbuch4DEGB} and solve for the mass $M$. Doing so, we find
\begin{equation}\label{eq:bhh}
M_\mathrm{int} = \sqrt{\alpha} = M^\mathrm{BH}_\mathrm{min},
\end{equation}
or that the Buchdahl bound   asymptotically approaches the smallest black hole mass allowed by the theory.

These results imply that it is possible to have stable compact objects in 4DEGB whose radii are smaller than that of the GR Buchdahl bound $R \geq 9M/4$  or even that of the Schwarzschild radius $R = 2M$.  We shall demonstrate below that this situation is indeed realized, and that quark stars can have radii that are arbitrarily close to that of a black hole of the same mass.

\subsection{Results for $\lambda \geq 0$}

\begin{figure*}
        \subfloat[\label{fig:tov a0}]{
        \includegraphics[width=7.6cm]{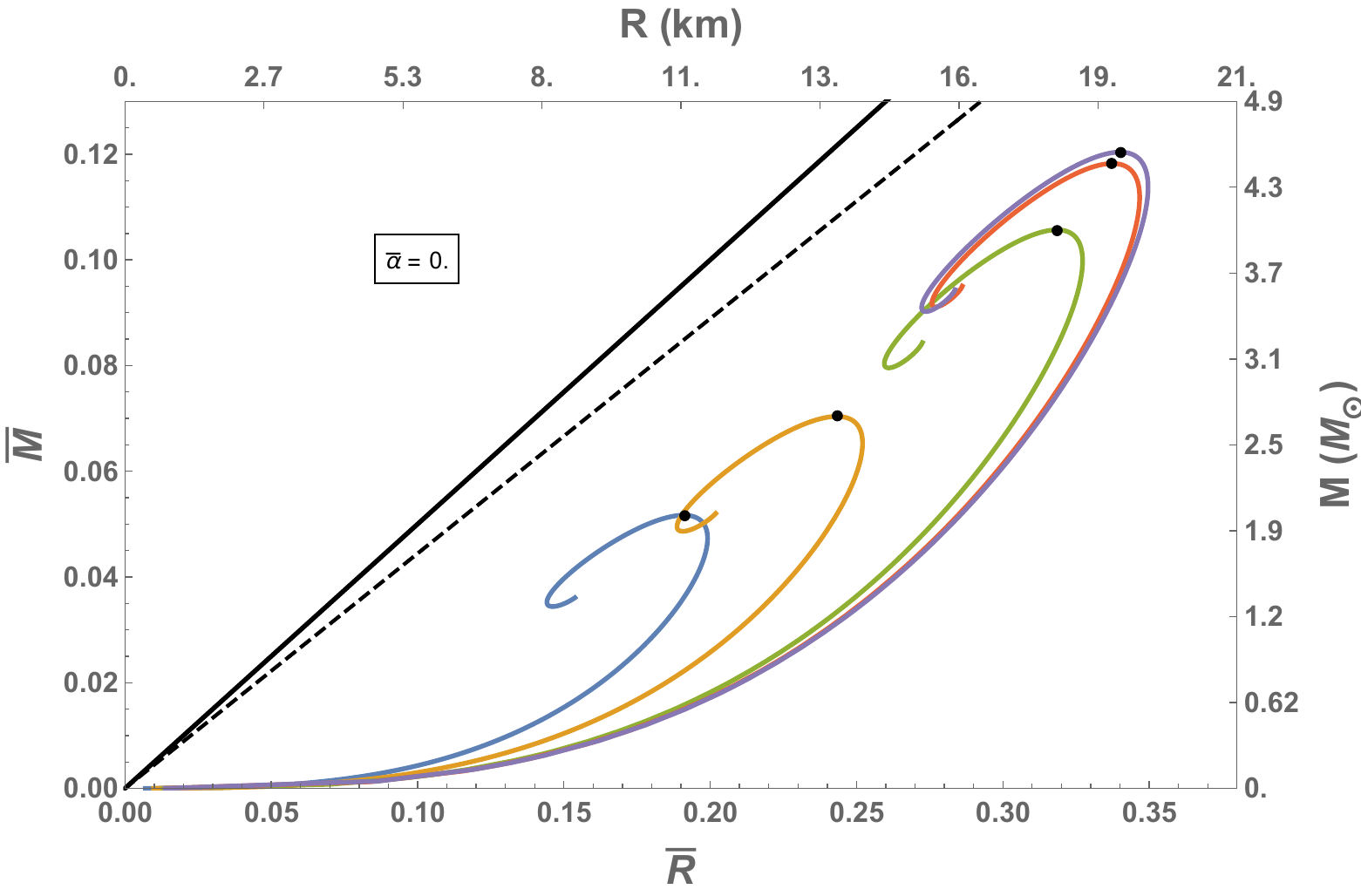}
        }\hfill
        \subfloat[\label{fig:mpc a0}]{
        \includegraphics[width=7.6cm]{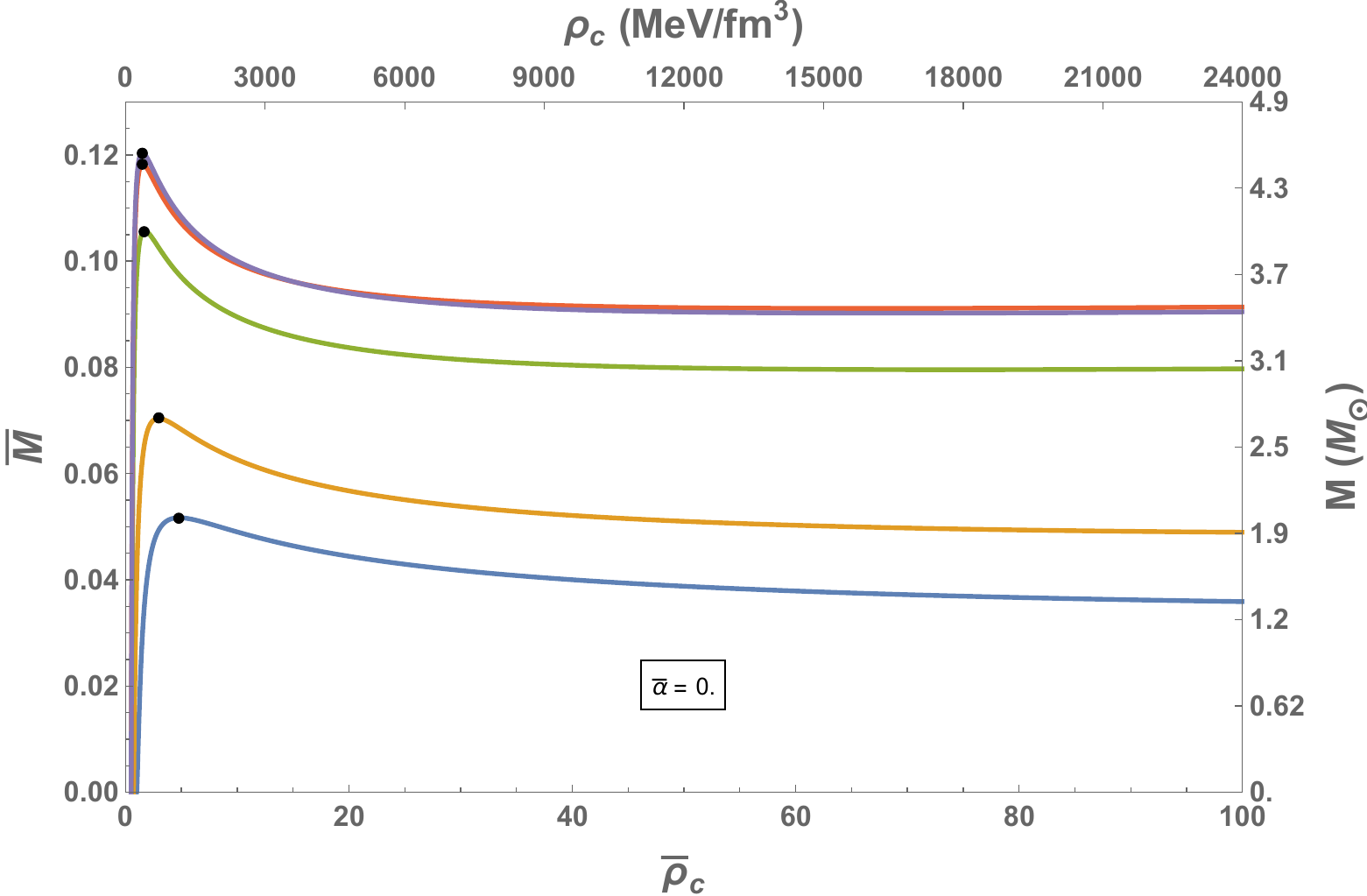}
        }

        \subfloat[\label{fig:tov apt01}]{
        \includegraphics[width=7.6cm]{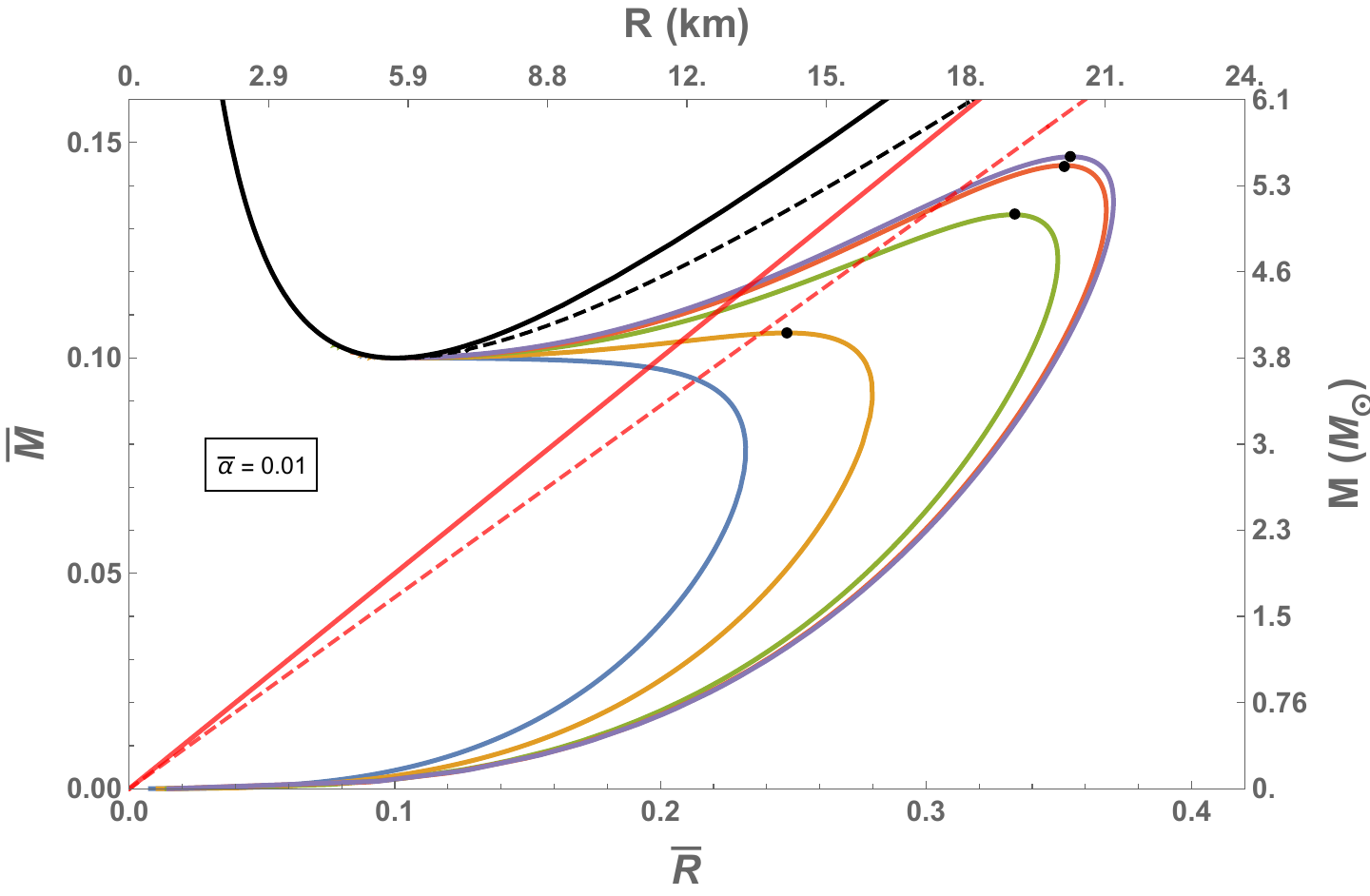}
        }\hfill
        \subfloat[\label{fig:mpc apt01}]{
        \includegraphics[width=7.6cm]{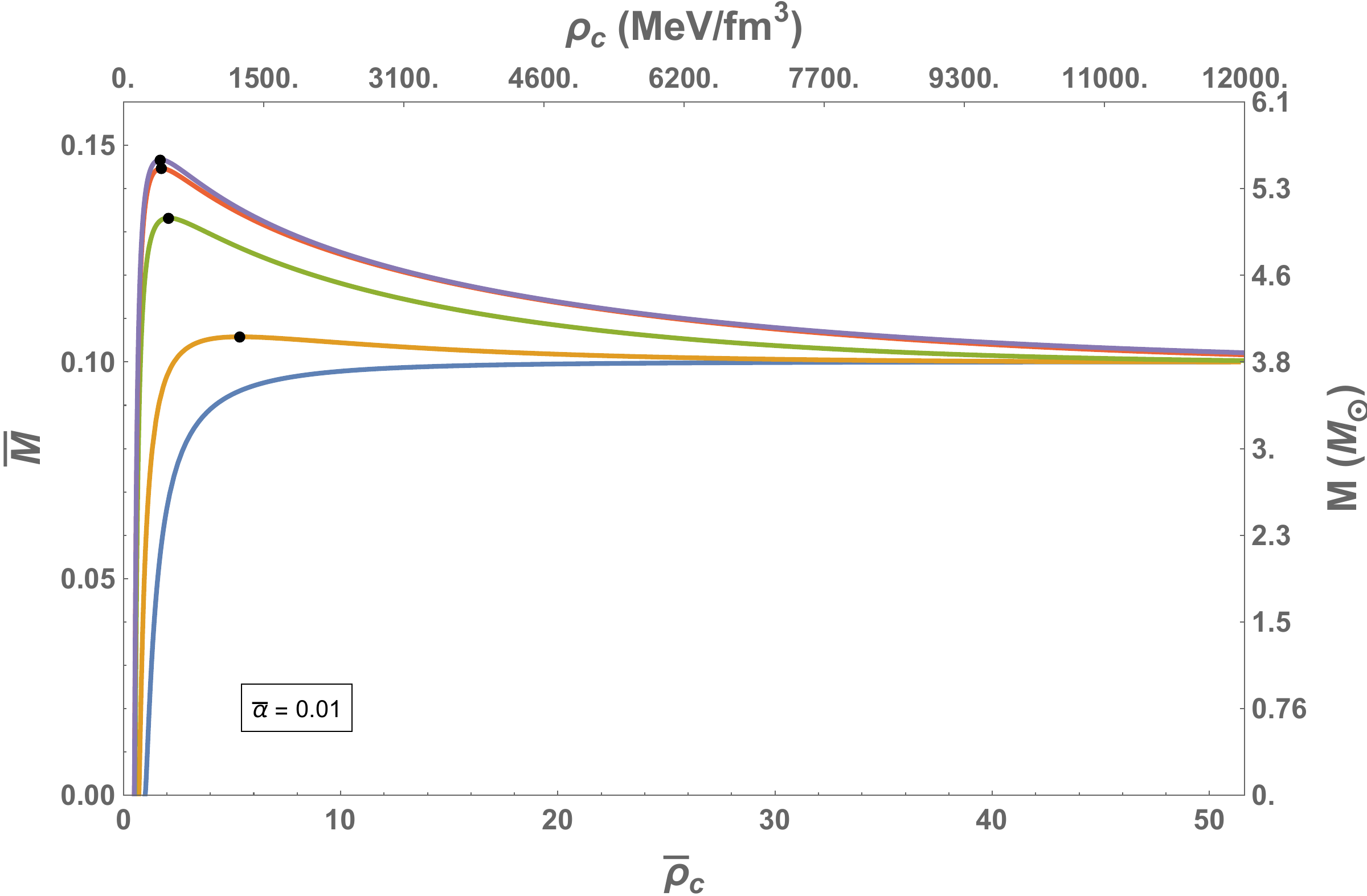}
        }

        \subfloat[\label{fig:tov apt05}]{
        \includegraphics[width=7.6cm]{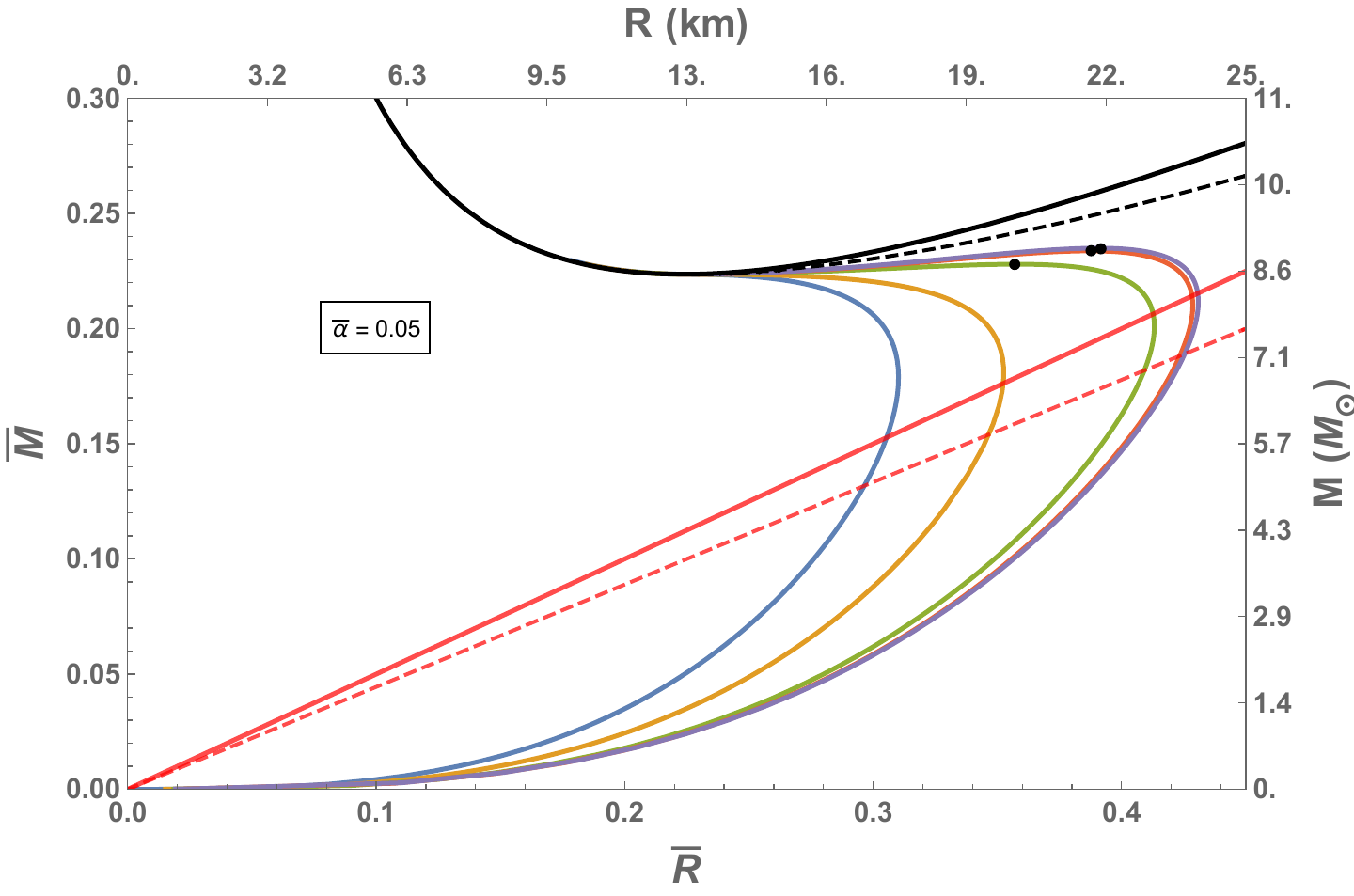}
        }\hfill
        \subfloat[\label{fig:mpc apt05}]{
        \includegraphics[width=7.6cm]{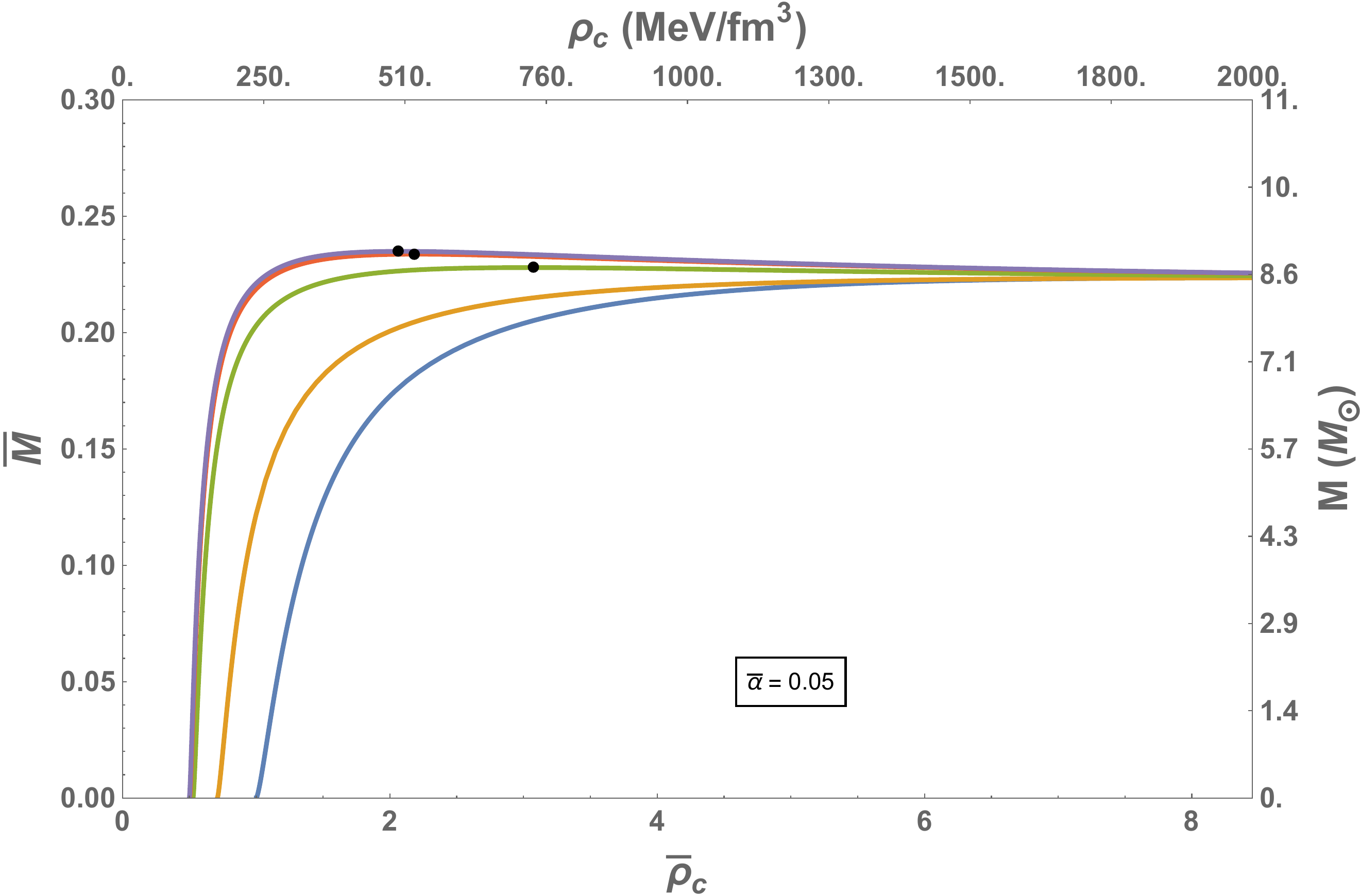}
        }
        
	\caption[]{Mass vs. radius/central density curves for unified interacting quark stars when $\lambda>0$. Each plot shows results for a unique value of the 4DEGB coupling. In order from blue to purple curves, the $\bar{\lambda}$ values considered are 0, 0.5, 10, 100, $\infty$ respectively. The solid and dashed black lines correspond to the 4DEGB equivalent of the Schwarzschild and Buchdahl limits, respectively (with their red counterparts marking the equivalent bounds in GR). In general a larger $\alpha$ and/or $\lambda$ tends to increase the mass and radius of a solution.  \label{fig:mr pos lambda}}
\end{figure*}

\begin{figure*}

        \subfloat[\label{fig:tov apt1}]{
        \includegraphics[width=7.6cm]{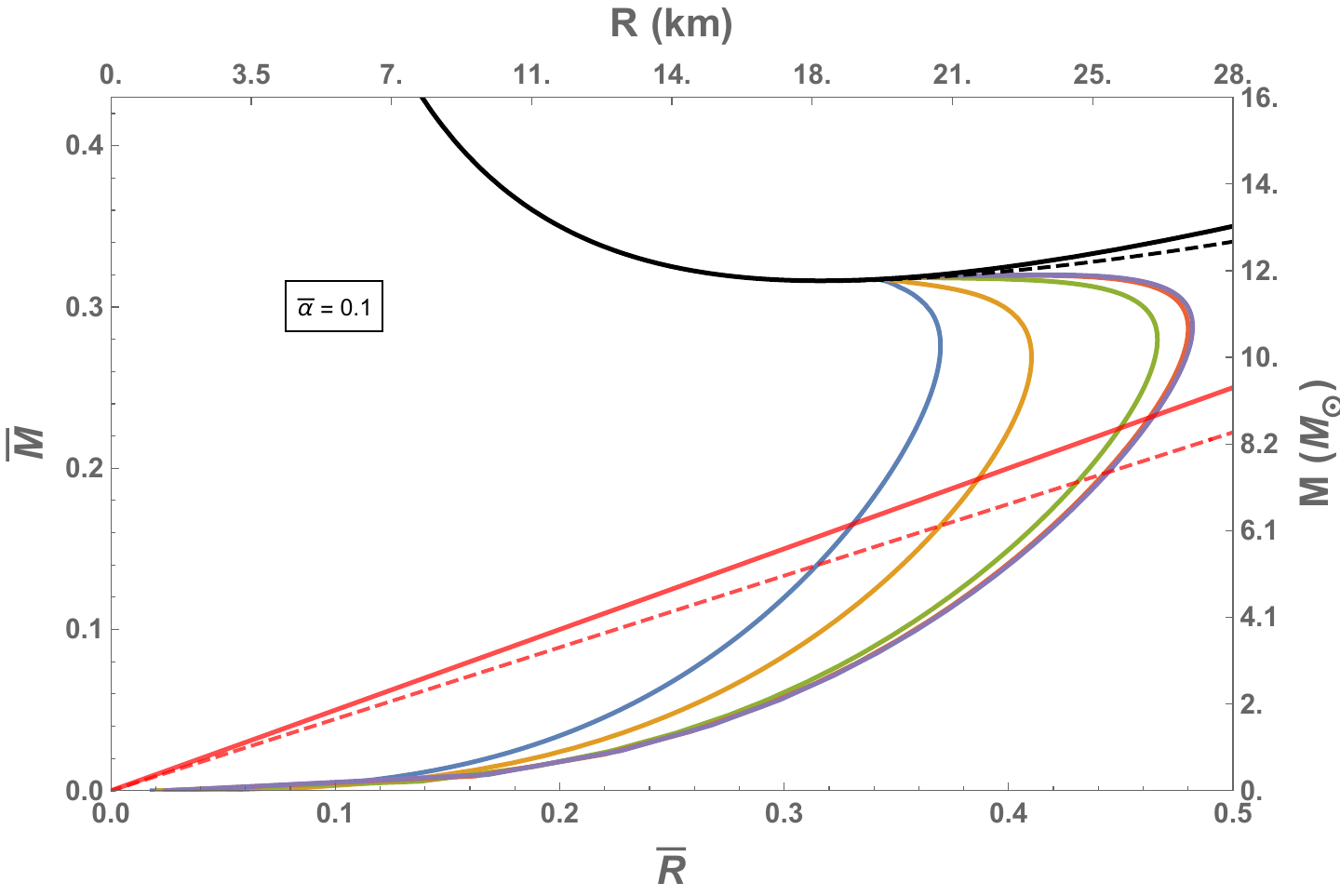}
        }\hfill
        \subfloat[\label{fig:mpc apt1}]{
        \includegraphics[width=7.6cm]{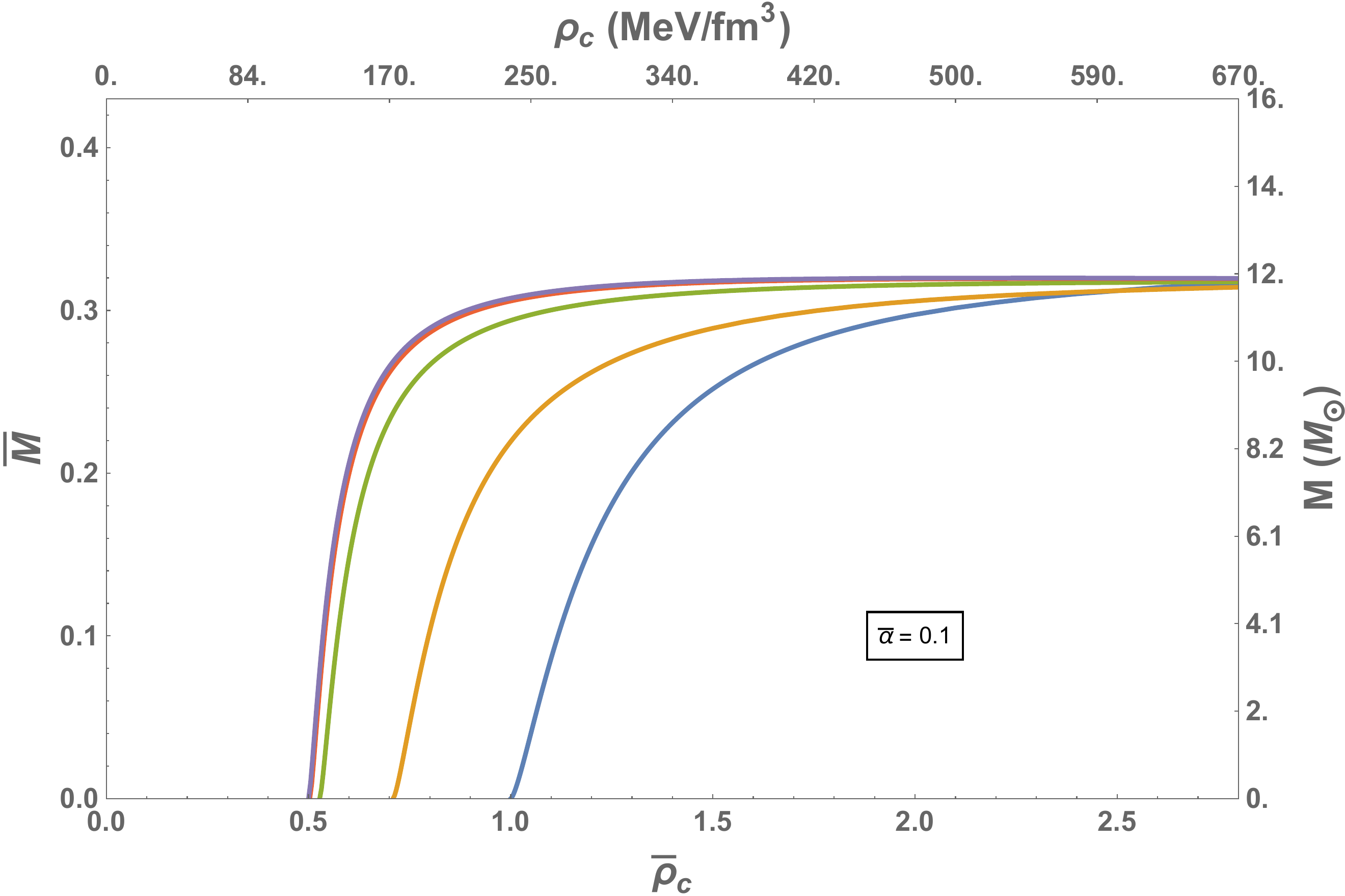}
        }

        \subfloat[\label{fig:tov apt4}]{
        \includegraphics[width=7.6cm]{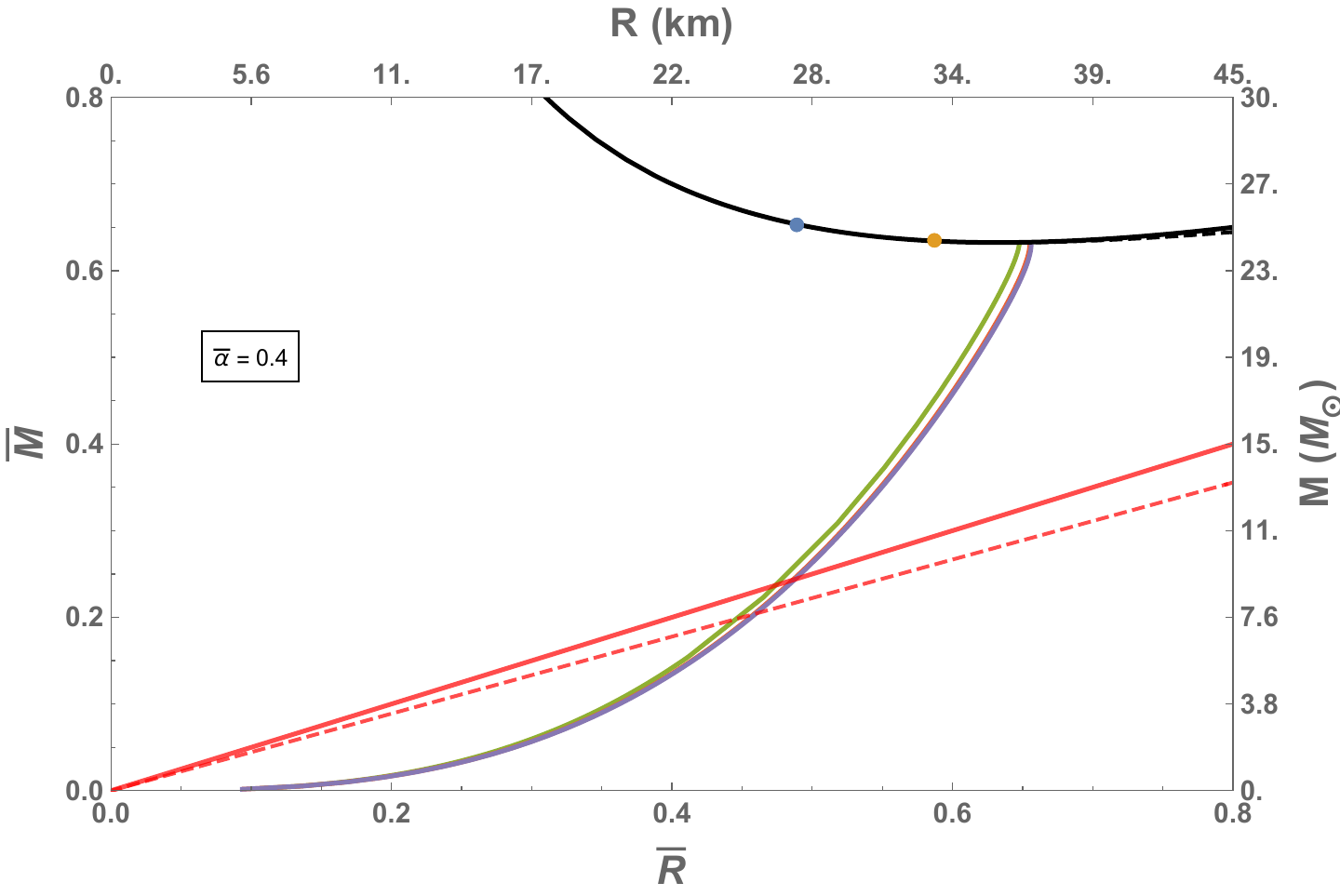}
        }\hfill
        \subfloat[\label{fig:mpc apt4}]{
        \includegraphics[width=7.6cm]{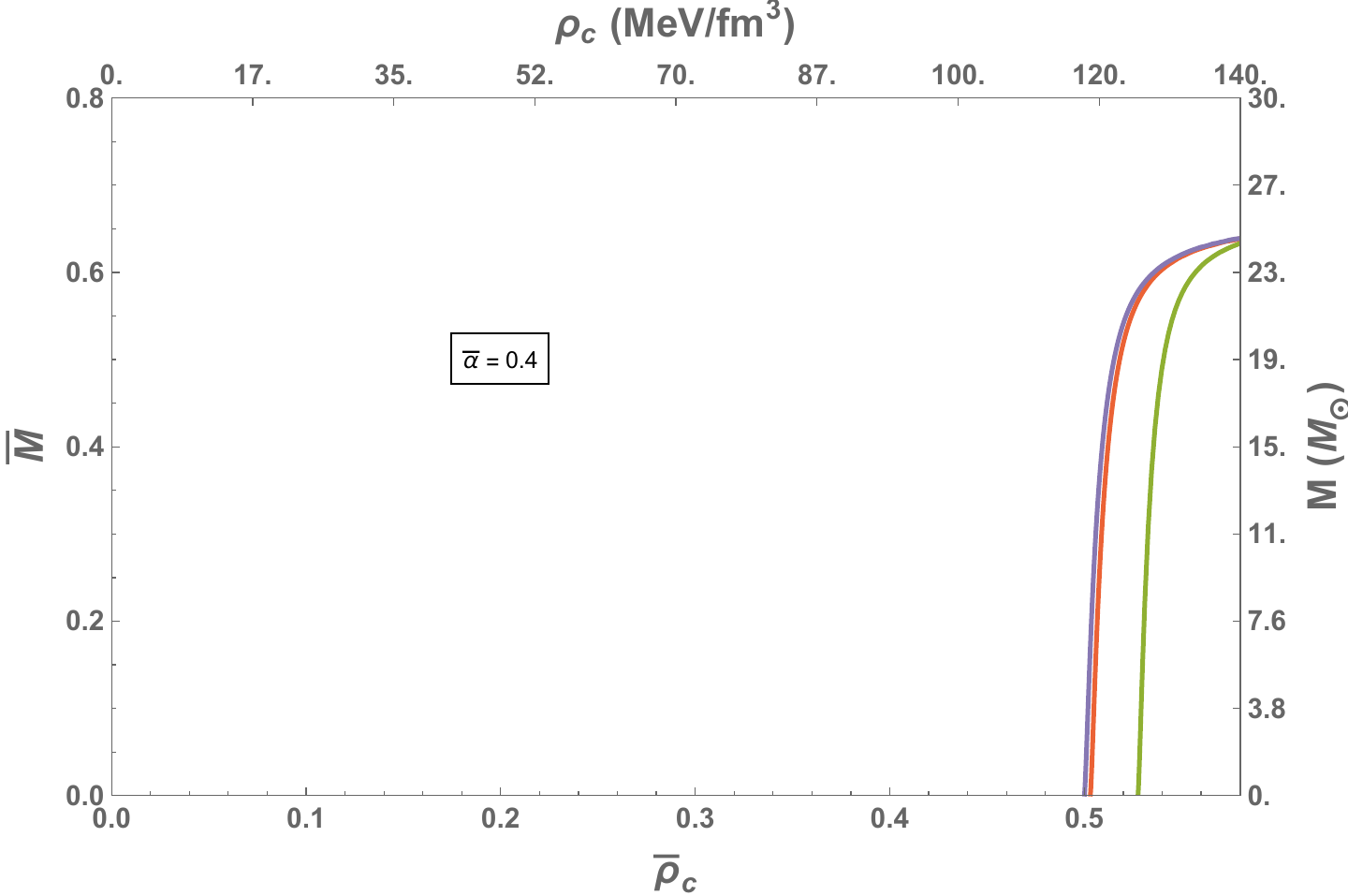}
        }

        \subfloat[\label{fig:tov a1}]{
        \includegraphics[width=7.6cm]{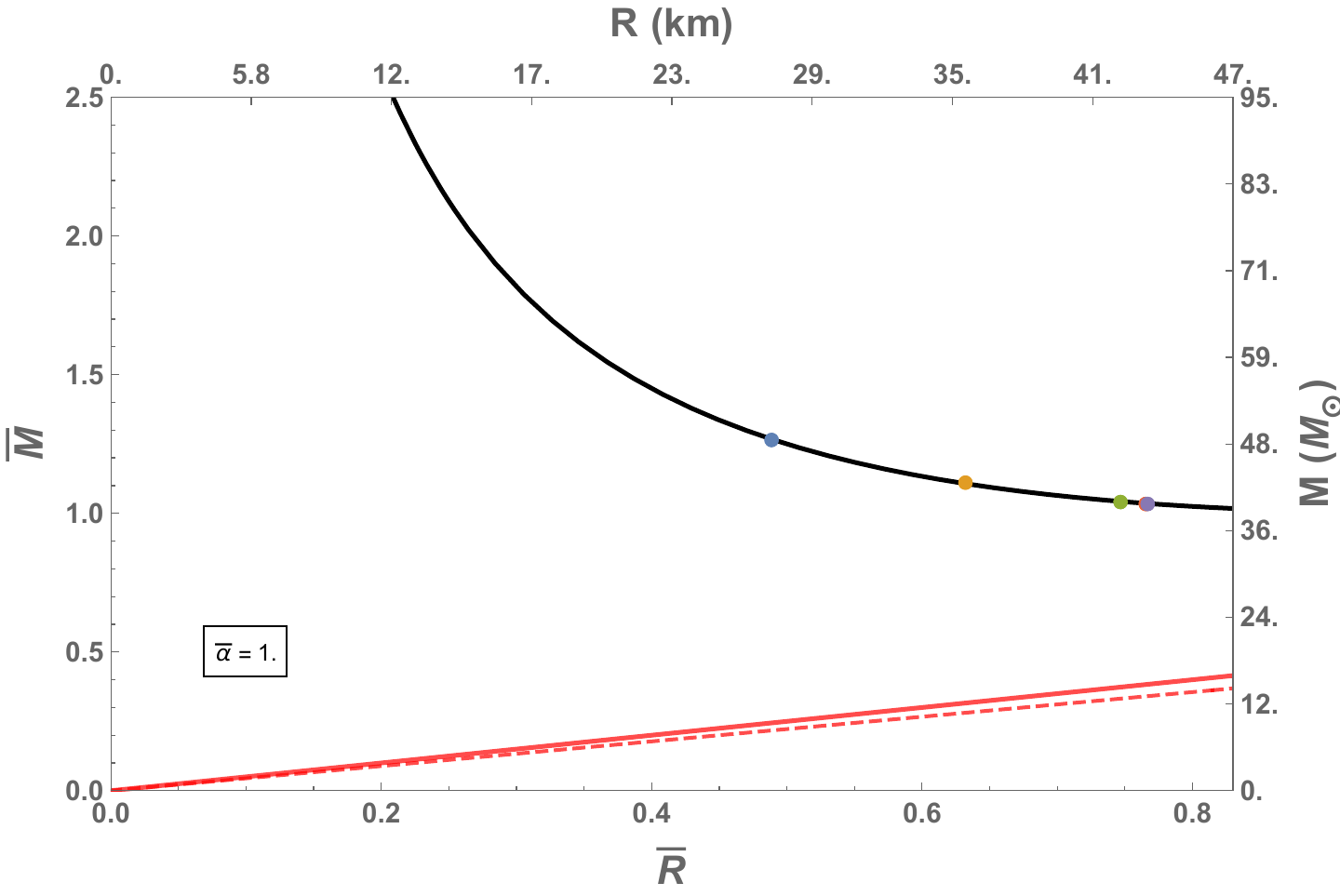}
        }\hfill
        \subfloat[\label{fig:mpc a1}]{
        \includegraphics[width=7.6cm]{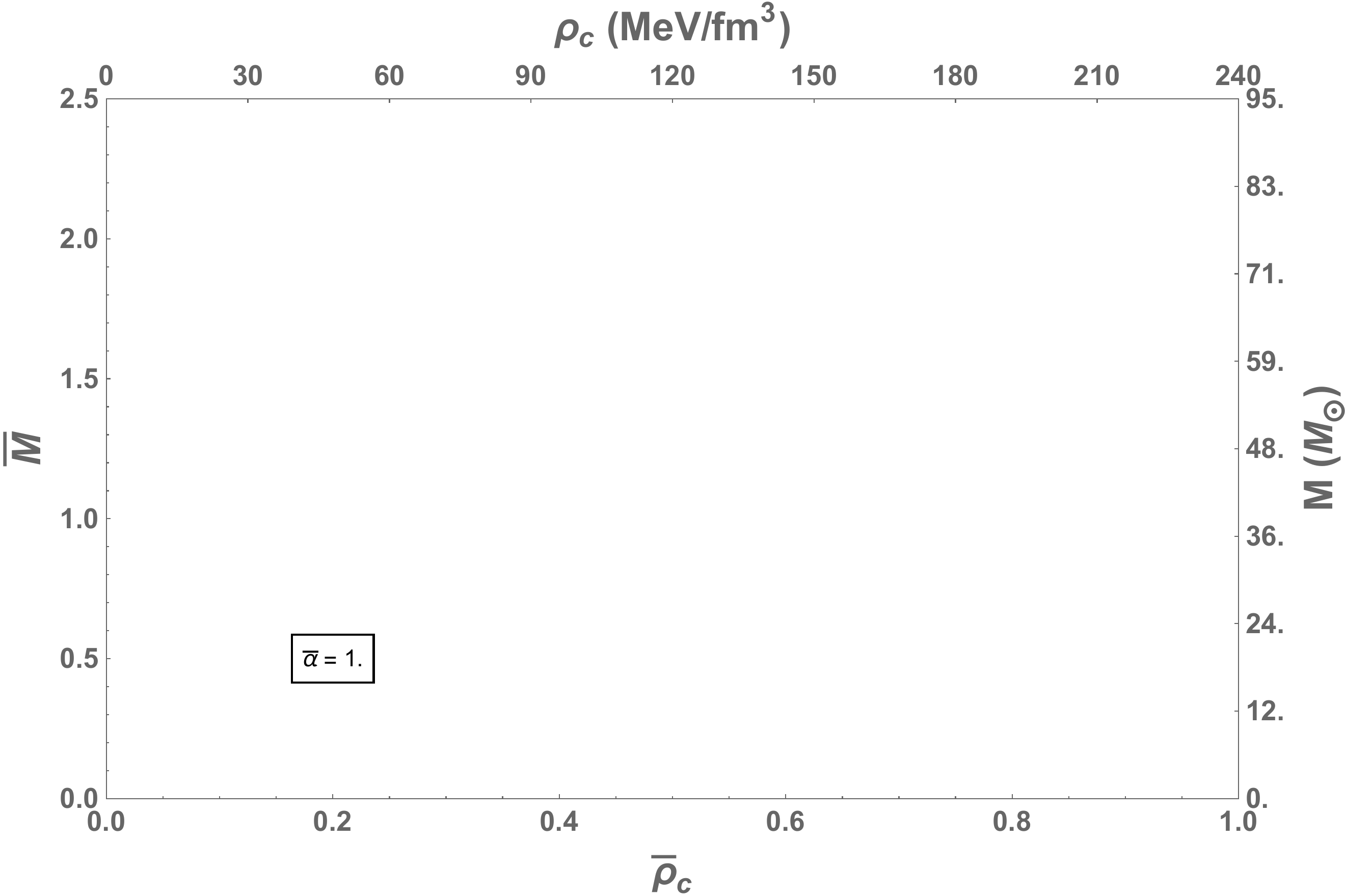}
        }

	\caption[]{Additional mass vs. radius/central density curves for unified interacting quark stars when $\lambda>0$. In order from blue to purple curves, the $\bar{\lambda}$ values considered are 0, 0.5, 10, 100, $\infty$ respectively. The solid and dashed black lines correspond to the 4DEGB equivalent of the Schwarzschild and Buchdahl limits, respectively (with their red counterparts marking the equivalent bounds in GR). In general a larger $\alpha$ and/or $\lambda$ tends to increase the mass and radius of a solution. At $\alpha/\lambda$ combinations large enough for $p_\mathrm{crit} > 0$ we observe solutions which \textit{start} at the black hole horizon (as shown in the lower two panels represented by coloured dots). These solutions are unstable for all choices of central pressure.\label{fig:mpc pos lambda}}
\end{figure*}

Here we present the solutions to the 4DEGB TOV equations for a positive coupling constant $\lambda \geq 0$. This is equivalent to the condition
\begin{equation}\label{eq:poslamcondition}
    \xi_{2 a} \Delta^2>\xi_{2 b} m_s^2,
\end{equation}
and clearly from \eqref{lameff} as  $\lambda$ increases in magnitude, so does $\bar{\lambda}$. Our results are illustrated for a range of values of
$\bar{\alpha}$ and $\bar{\lambda}$ in Figures~\ref{fig:mr pos lambda} and ~\ref{fig:mpc pos lambda}.

We find that in general, both increasing $\bar{\alpha}$ and increasing $\bar{\lambda}$ expand the curves to larger mass and radii for the same choice of central pressure, yielding larger maximum mass points. 
This is due to gravitational attraction in 4DEGB being increasingly weaker as $\alpha >0$ increases, as well as a larger $\lambda$ mapping to a stiffer equation of state as a consequence of the strong interaction effects. We see that for any given $\alpha$, the hook-shaped $M$~vs.~$R$ curves move upward and to the right as $\lambda$ increases. Likewise, for any given $\lambda$, the curves again move upward and rightward, with the upper part more rapidly asymptoting to the Buchdahl/horizon limit as $\alpha$ increases.

\begin{figure}
    \subfloat[\label{fig:tov apt001}]{
        \includegraphics[width=6.8cm]{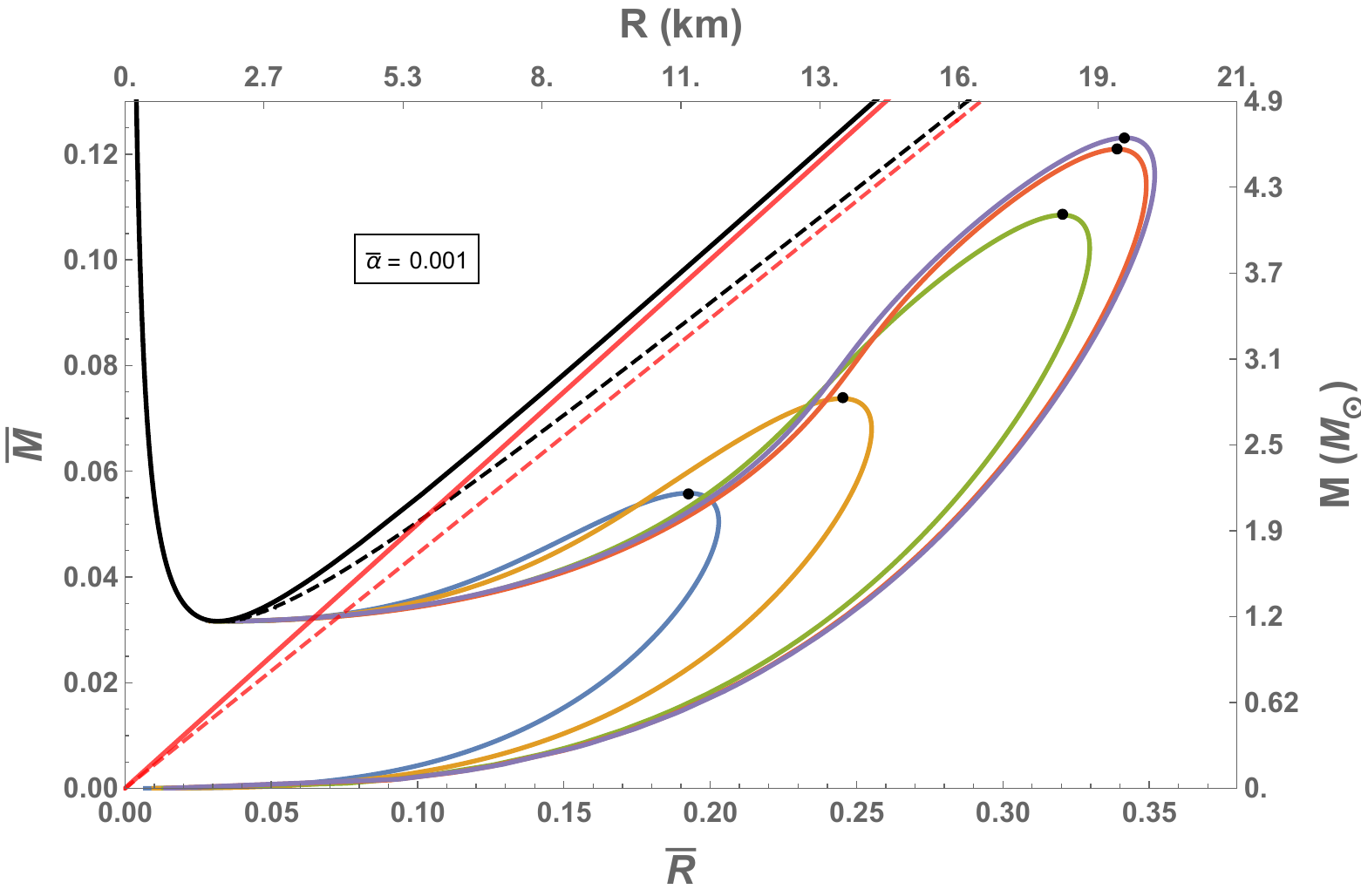}
        }\hfill
    \subfloat[\label{fig:mpc apt001}]{
        \includegraphics[width=6.8cm]{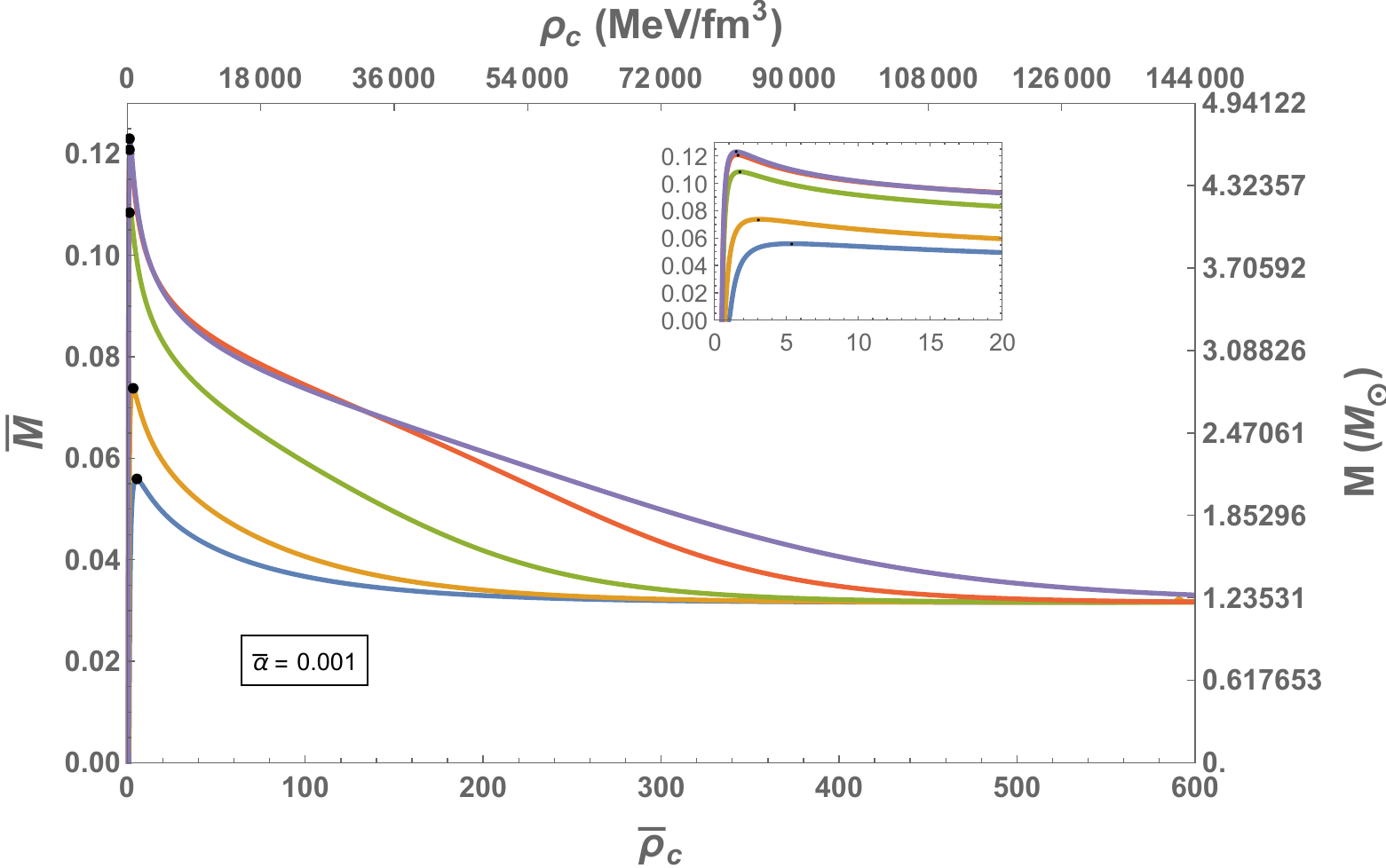}
        }\hfill
    \subfloat[\label{legend}]{
    \includegraphics[width=1.6cm]{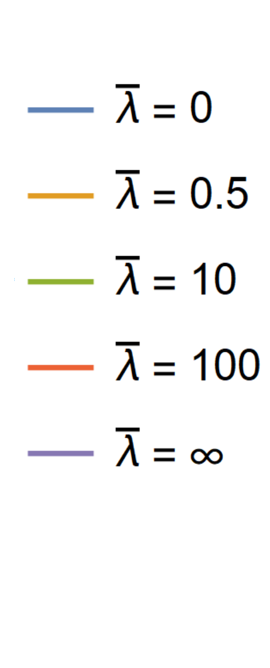}
        }

        \caption[]{Mass vs. radius and central density for $\bar{\alpha} = .001$ when $\lambda>0$. We see that even for very small alpha there exists a range of solutions yielding quark stars smaller than the GR Buchdahl bound (dashed red) and the $2M$ Schwarzschild radius (solid red); the corresponding bounds for 4DEGB are in black.
        \label{fig:alpha small}}
\end{figure}

It is quite interesting to note that even for small $\alpha$, with a high enough central pressure there are stable solutions that are smaller in radius than both the GR Buchdahl bound and the Schwarzschild black hole.
We illustrate this in Fig.~\ref{fig:alpha small}, which shows that, even for $\bar{\alpha}=0.001$ there is a narrow range for small mass, small radius stars that are within not only the GR Buchdahl bound, but also the $2M$ Schwarzschild radius. Similar results can be seen in the 2nd panel of Figure \ref{fig:tov apt0001 neglam} for $\bar{\alpha}=0.0001$ when $\lambda<0$. The stability of these objects remains an interesting subject for investigation.

Our results are commensurate with those of Banerjee et al. \cite{banerjee_2021_quark} in the limited overlapping regions 
of parameter space,
in that increasing the coupling strength to the higher curvature terms tends to increase the mass-radius profiles, in turn yielding a larger local maximal mass point. In our case not all curves have this local extrema, but in the small $\alpha$ and $\rho_c$ regime we see the same behaviour. In \cite{banerjee_2021_quark}, rather than varying coupling strength directly,  different values of the MIT bag constant are considered, with a \textit{smaller} $B_\mathrm{eff}$ corresponding to larger M-R profiles (which in our case is correlated with a larger positive interaction strength).


Furthermore, for sufficiently large 
$\alpha/\lambda$, no quark star solutions exist outside of the black hole horizon, since the pressure profiles smaller than criticality diverge to $\infty$ with no real roots. 
This in turn implies 
that there should be a well-defined central pressure/density at which $p(r) = p(0) = \mathrm{const}$, and below which quark star solutions do not exist. This value can be derived analytically (see Appendix \ref{sec:appendixcritpres} for a detailed derivation), and we find that below the critical central pressure:
\begin{equation}
\begin{aligned}
    p_\mathrm{crit} = \frac{1}{2 \pi  (9 \pi -32 \alpha  \lambda )^2} &\Bigg[256 \alpha  (2 \pi  \alpha +3) \lambda ^2+8 \pi  (2 \pi  \alpha  (8 \pi  \alpha +21)-9) \lambda +9 \pi ^3 (4 \pi  \alpha -3)\\
    &-\mathrm{sgn}(\lambda)6 \sqrt{\lambda  (32 \alpha  \lambda +\pi  (8 \pi  \alpha -3))^2 \left(16 (2 \pi  \alpha +1) \lambda +\pi ^2 (8 \pi  \alpha +3)\right)}\Bigg],
    \end{aligned}
\end{equation}
there are no real roots of $p(r)$ with which to define the star's surface.


This central pressure is never realized physically. We find that all
$M$~vs.~$R$ curves with $p_\mathrm{crit}>0$ begin at the 4DEGB black hole horizon and extend to values within the horizon, or in other words, in a disallowed region of parameter space. We illustrate this in Figure \ref{fig:tov apt4}, where we notice that the $M$~vs.~$R$ curves with $p_\mathrm{crit}>0$ begin at the 4DEGB black hole horizon (as coloured points). The first threshold of interest is to consider is the smallest value of $\bar{\alpha}$ at $\bar{\lambda}=0$ for which none of the solutions lie below the Buchdahl threshold -- for such parameter combinations no physical quark star solutions exist.

To investigate this further, consider the pressure profile of a star with a central pressure just past criticality (ie. $p(0) = p_\mathrm{crit} + \delta$ where $\delta<<1$). As shown in Appendix \ref{sec:appendixcritpres}, the pressure profile is well-approximated by a step function times a constant, allowing us to approximately solve \eqref{mdiff} analytically:
\begin{equation}
    \bar{m}(\bar{R})^\mathrm{min} = \bar{M}^\mathrm{min} \approx 4 \pi \int_0^{\bar{R}} \Theta(\bar{r}-\bar{R}) \bar{\rho}(0) \bar{r}^2 d\bar{r} = \frac{4 \pi}{3}\bar{R}^3 \bar{\rho}(0).
\end{equation}
Then for the special case of vanishing interaction strength,
\begin{equation}
    \bar{M}_{\lambda \to 0}^\mathrm{min} \approx \frac{4 \pi}{3}\bar{R}^3 (3 \bar{p}(0) + 1).
\end{equation}
Since $p(0) \approx p_\mathrm{crit}$ and $p_\mathrm{crit}^{\lambda \to 0} = \frac{2 \pi \bar{\alpha}}{9} - \frac{1}{6}$, this can be further simplified into the following form:
\begin{equation}\label{eq:mminlam0}
\bar{M}_{\lambda \to 0}^\mathrm{min} \approx \frac{2}{3} \pi  (\frac{4 \pi}{3}  \bar{\alpha} + 1) \bar{R}^3.
\end{equation}
Inserting this minimal mass solution into \eqref{pdiff} and finding where $\bar{p}'(\bar{r})$ diverges, we obtain the corresponding radius 
\begin{equation}\label{eq:rminlam0}
    \bar{R}_{\lambda \to 0}^\mathrm{min} \approx \sqrt{\frac{3}{4\pi}}
\end{equation}
of this minimal mass point, which is (interestingly) independent of $\bar{\alpha}$.
Consequently 
\begin{equation}
 {\bar{M}_{\lambda \to 0}^\mathrm{min}} \approx \frac{3 + 4 \pi  \bar{\alpha}}{4 \sqrt{3 \pi }}.
\end{equation}
Inserting this  into the full 4DEGB Buchdahl bound and solving for $\bar{\alpha}$, we find that 
\eqref{eq:fullbuch4DEGB} 
is saturated when $\bar{\alpha} = \frac{3}{4 \pi}$, which consequently is also the solution to $\lim_{\lambda \to 0} p_\mathrm{crit}(\bar{\alpha}) = 0$ (the smallest $\bar{\alpha}$ with a non-negative critical pressure in the limit of vanishing QM interaction), implying that $\bar{M}_{\lambda \to 0}^\mathrm{min}|_{\bar{\alpha}=\frac{3}{4 \pi}} = \bar{M}_\mathrm{int}$. 

For a more general choice of $\bar{\alpha}$ it is easy to see that by inserting  \eqref{eq:rminlam0} into the 4DEGB black hole mass $M_{BH} = \frac{\alpha + R_H^2}{2R_H}$ that $M_{BH}(\bar{R}_{\lambda \to 0}^\mathrm{min}) = \bar{M}_{\lambda \to 0}^\mathrm{min}$. In other words, all $\bar{\lambda} = 0$ solution curves with non-zero critical pressure must \textit{start} at the black hole horizon (or the Buchdahl bound/horizon intersection, in the case where $\bar{\alpha}=\frac{3}{4 \pi}$), and thus none of these solutions are stable. This can also be checked numerically by considering $\bar{\alpha} = \frac{3}{4 \pi} + \delta$ (where $\delta<<1$) and plotting the lowest pressure solution possible, which indeed intersects the minimal mass point of the Buchdahl curve.

A similar analysis can be repeated in the limit $\lambda \to \infty$. Doing so we find an equivalent expression for the mass of the threshold point:
\begin{equation}
\bar{M}_{\lambda \to \infty}^\mathrm{min} \approx \frac{\left(1 + 2 \pi  \bar{\alpha} -\sqrt{2 \pi  \bar{\alpha} +1}\right)}{2 \bar{\alpha} } \bar{R}^3
\end{equation}
and radius:
\begin{equation}
    \bar{R}_{\lambda \to \infty}^\mathrm{min} \approx  \sqrt{\frac{2 \bar{\alpha} }{\sqrt{8 \pi  \bar{\alpha} -4 \sqrt{2 \pi  \bar{\alpha} +1}+5}-1}},
\end{equation}
the latter of which now has an explicit $\bar{\alpha}$ dependence. Combining the two we find
\begin{equation}\label{eq:mminlaminf}
\bar{M}_{\lambda \to \infty}^\mathrm{min} \approx \sqrt{2\bar{\alpha} } \left(2 \pi  \bar{\alpha} + 1 -\sqrt{2 \pi  \bar{\alpha} +1}\right) \left(\frac{1}{\sqrt{8 \pi  \bar{\alpha} -4 \sqrt{2 \pi  \bar{\alpha} +1}+5}-1}\right)^{3/2}.
\end{equation}

\begin{figure}
    \centering
    \includegraphics[width=12cm]{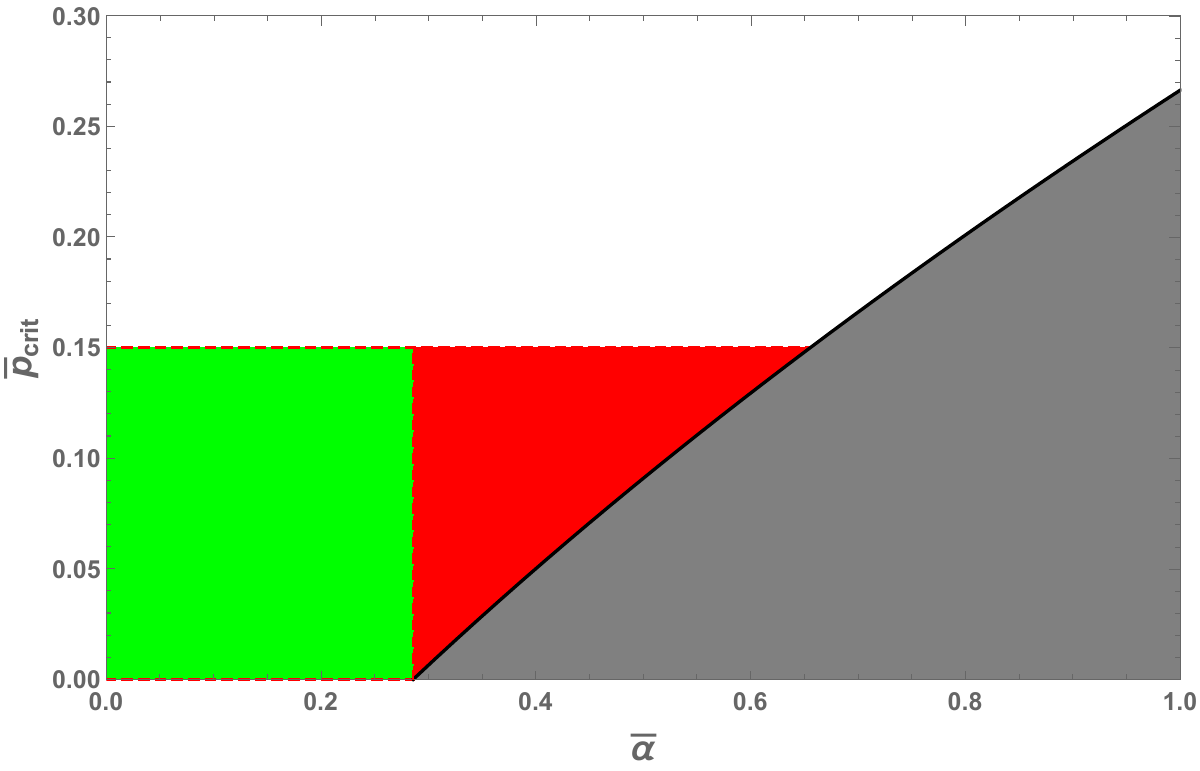}
    \caption{The solid black line 
    (bordering the gray region) 
    shows $p_\mathrm{crit}(\alpha)$ for a test value of $\bar{\lambda} = 0.1$ (and $\lambda>0$). For a fixed central pressure of $\bar{p}_0 = $ 0.15 the parameter space has three distinct regions. From $\bar{\alpha} = 0$ to the value that satisfies $p_\mathrm{crit}(\alpha) = 0$ (green) there exist
    valid, physical solutions
    at fixed $p_0$. As $\bar{\alpha}$ increases past this
    the red region becomes manifest, having  parameter combinations with $p_0>p_\mathrm{crit}$, and thus with unphysical radii inside the black hole horizon.
    the gray region has parameter combinations with $p_0<p_\mathrm{crit}$ (and thus radial pressure profiles that diverge to $\infty$).  The boundary between the gray and red regions corresponds with solutions lying directly at the black hole horizon - these are included in our M-R plots as coloured points.}
    \label{fig:pcritregions}
\end{figure}

As before, we  insert this mass/radius pair into the full 4DEGB Buchdahl bound \eqref{eq:fullbuch4DEGB}, and find that the expression is saturated when $\bar{\alpha} = \frac{3}{2\pi}$, once again the solution to $\lim_{\lambda \to \infty} p_\mathrm{crit}(\bar{\alpha}) = 0$. Inserting $\bar{\alpha} = \frac{3}{2\pi}$ in to \eqref{eq:mminlaminf} we confirm once again that $\bar{M}_{\lambda \to \infty}^\mathrm{min}|_{\bar{\alpha} = \frac{3}{2\pi}} = M_\mathrm{int}$, implying that solutions with critical pressure will \textit{start} at the intersection point of the 4DEGB Buchdahl bound and black hole horizon, and are thus always unphysical. For an arbitrary $\bar{\alpha}>\frac{3}{2\pi}$ it is easy to show that again $M_{BH}(\bar{R}_{\lambda \to \infty}^\mathrm{min}) = \bar{M}_{\lambda \to \infty}^\mathrm{min}$, indicating that the lowest pressure solution intersects the black hole horizon. As with the non-interacting case, the $\bar{\lambda} \to \infty$ solutions are also unstable for any central pressure above the critical value, and thus for any $\bar{\alpha}/\bar{\lambda}$ combination with a positive critical pressure. These results are  manifest in Figures \ref{fig:mr pos lambda} and  \ref{fig:mpc pos lambda} as points lying along the solid black line.  The above analysis also breaks the parameter space up into three different regions, which can be seen in Figure \ref{fig:pcritregions}. 

\begin{figure}
    \subfloat[\label{fig:tov lam0}]{
        \includegraphics[width=7.6cm]{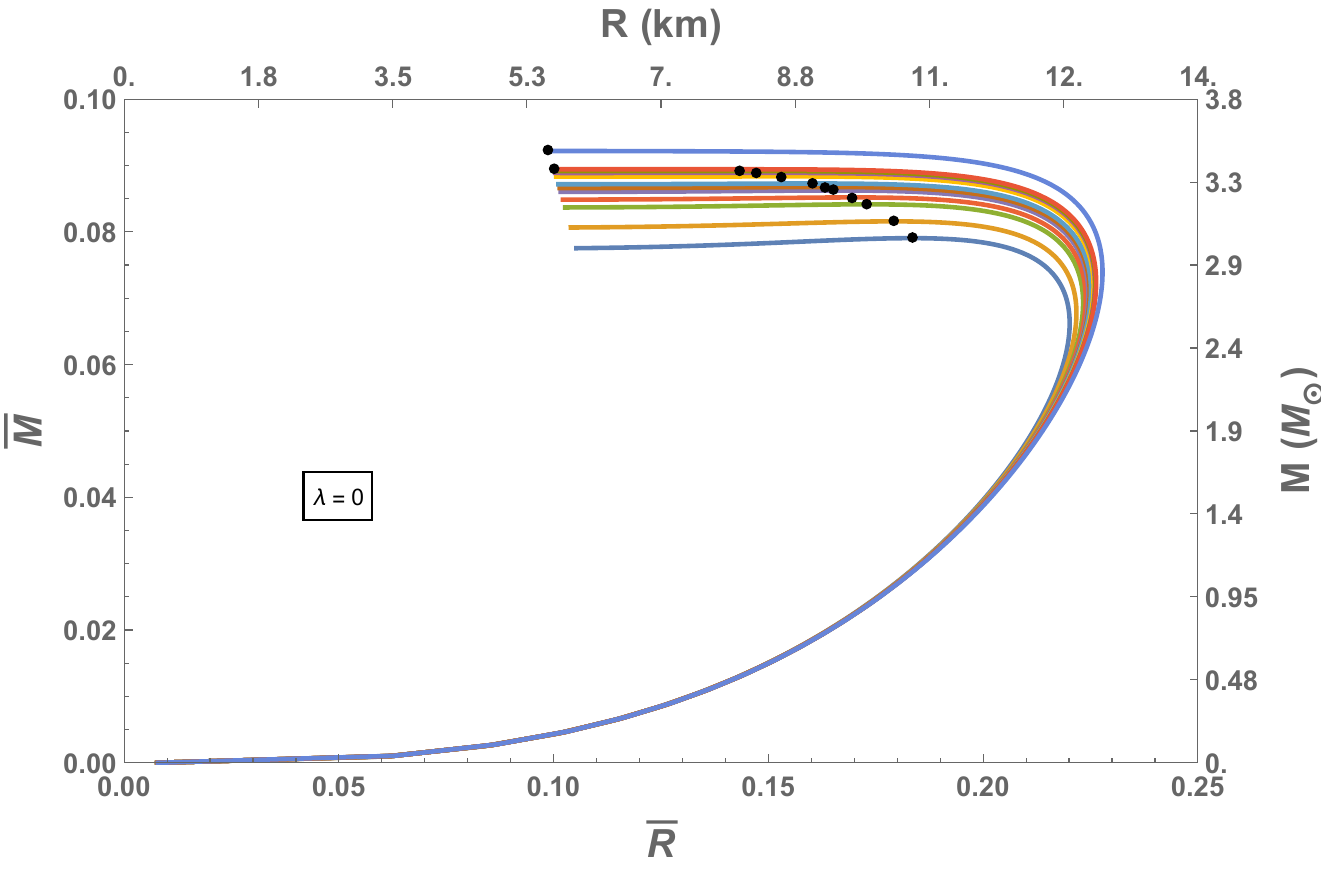}
        }\hfill
    \subfloat[\label{fig:mpc lam0}]{
        \includegraphics[width=7.6cm]{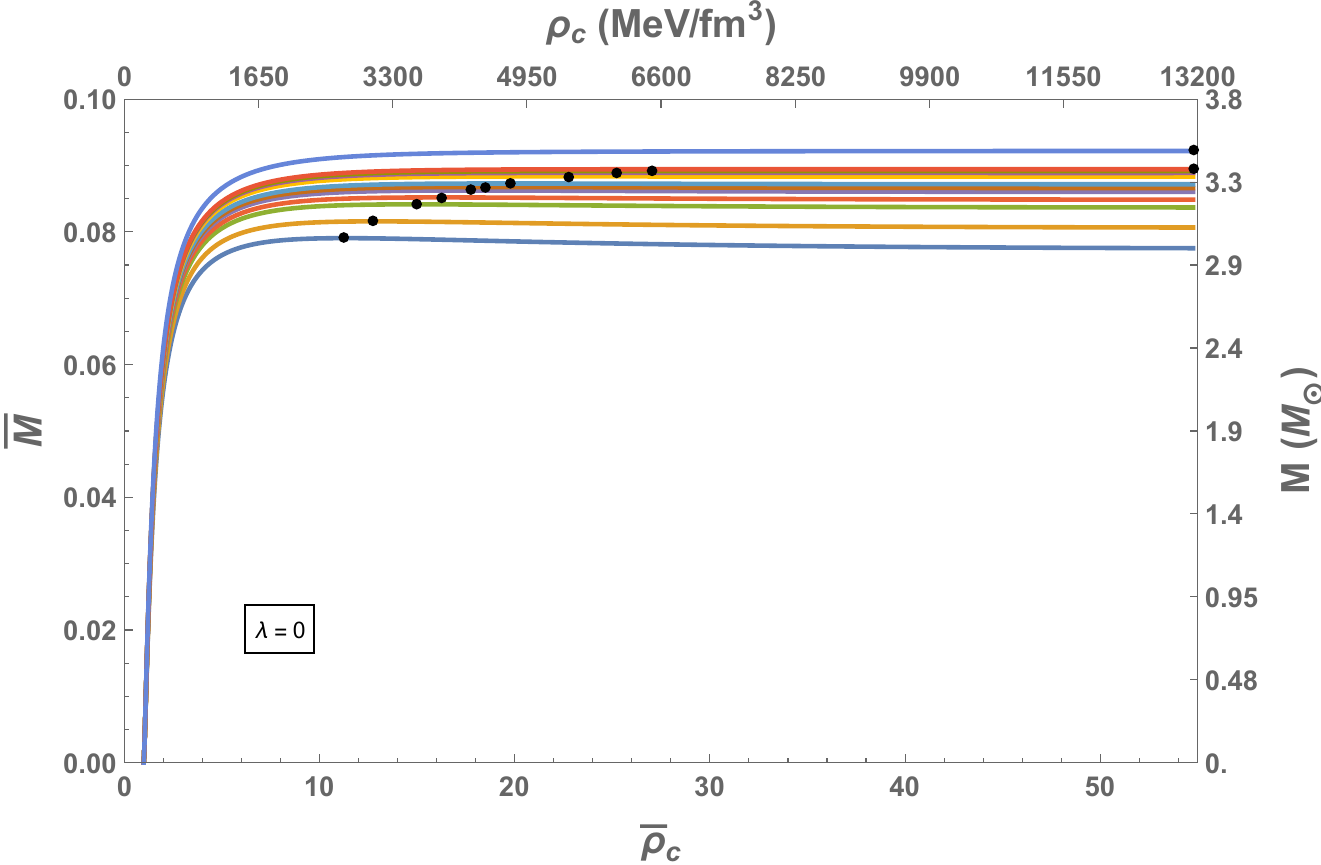}
        }

        \caption[]{Mass vs. radius and central density where the interaction strength is fixed at 0, and $\bar{\alpha} =$ 0.006, 0.0065, 0.007, 0.0072, 0.0074, 0.0075, 0.0076, 0.0078, 0.0079, 0.00795, 0.008, 0.0085 (lower $\bar{\alpha}$ corresponding to lower maximal mass points). The purpose of this plot is to show the progression of the maximum mass point as $\alpha$ changes slowly.
        \label{fig:alpha scan}}
\end{figure}

\subsection{Results for $\lambda \leq 0$}

\begin{figure*}
        \subfloat[\label{fig:tov a0 neglam}]{
        \includegraphics[width=7.6cm]{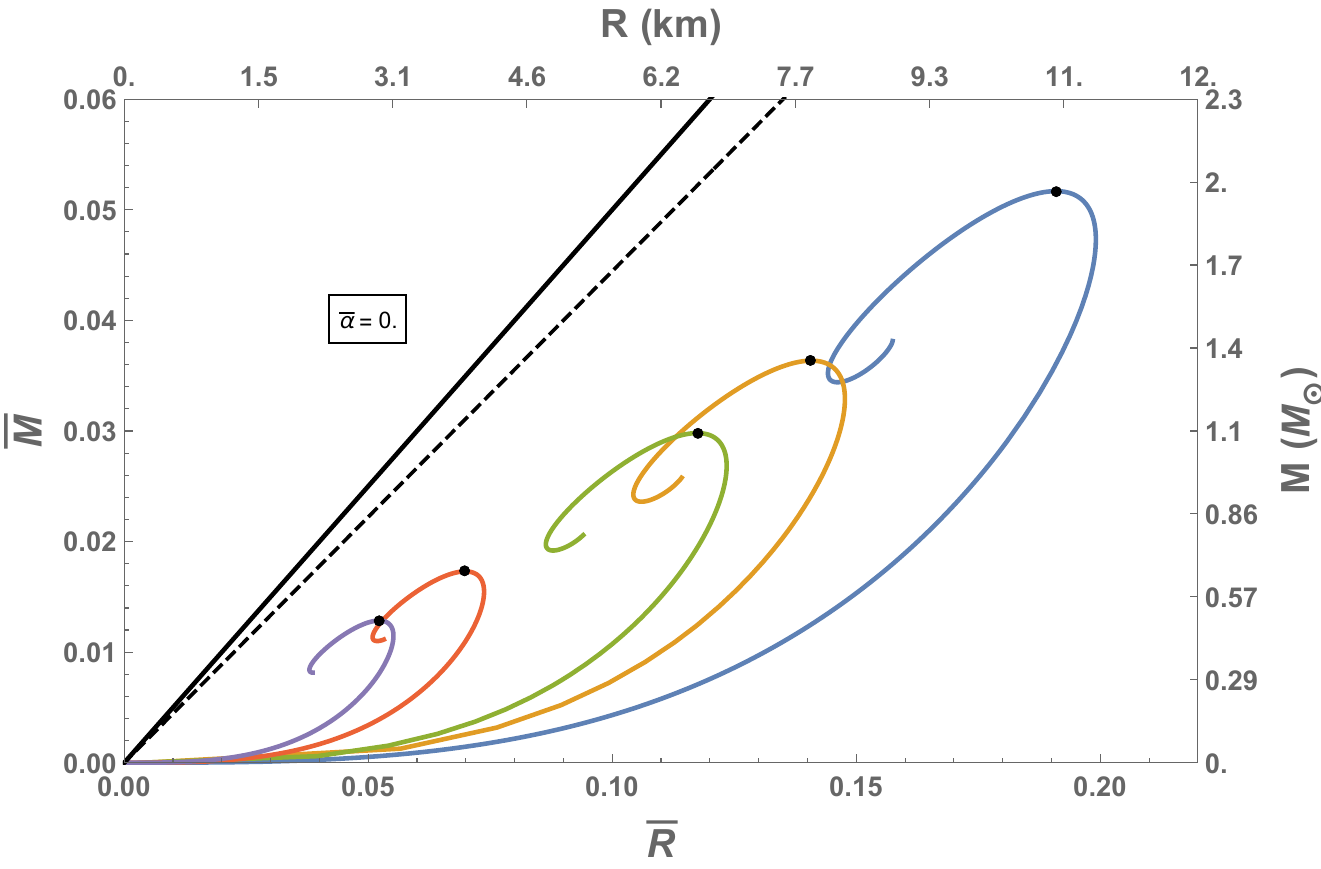}
        }\hfill
        \subfloat[\label{fig:mpc a0 neglam}]{
        \includegraphics[width=7.6cm]{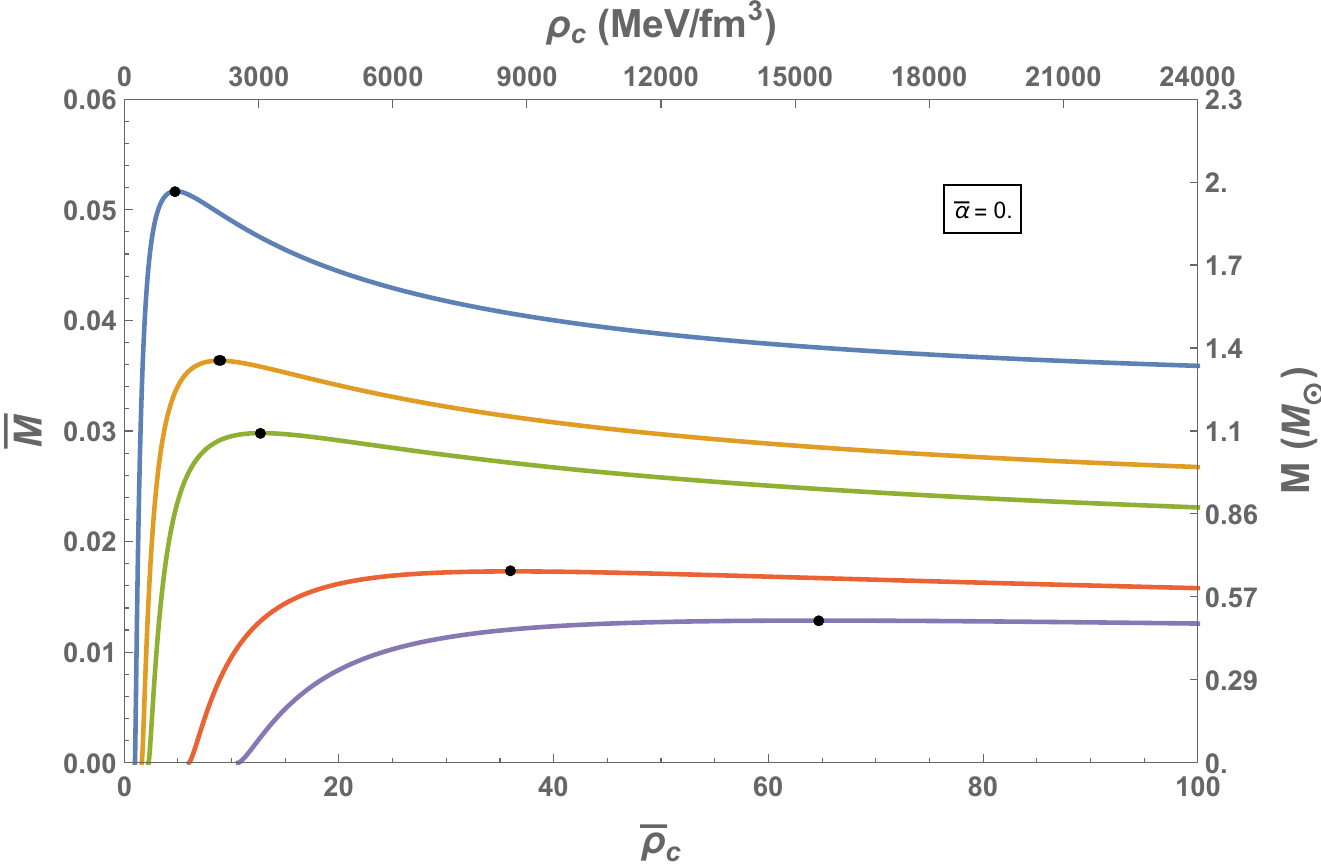}
        }

        \subfloat[\label{fig:tov apt0001 neglam}]{
        \includegraphics[width=7.6cm]{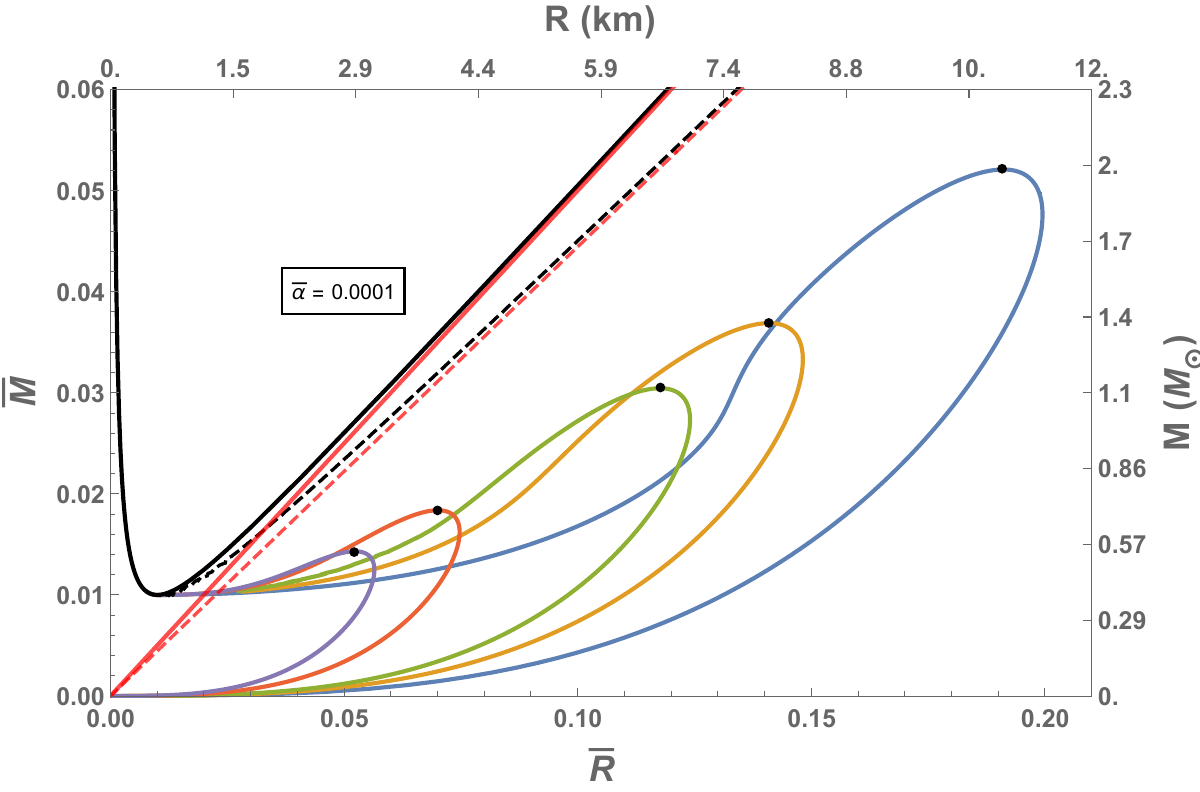}
        }\hfill
        \subfloat[\label{fig:mpc apt0001 neglam}]{
        \includegraphics[width=7.6cm]{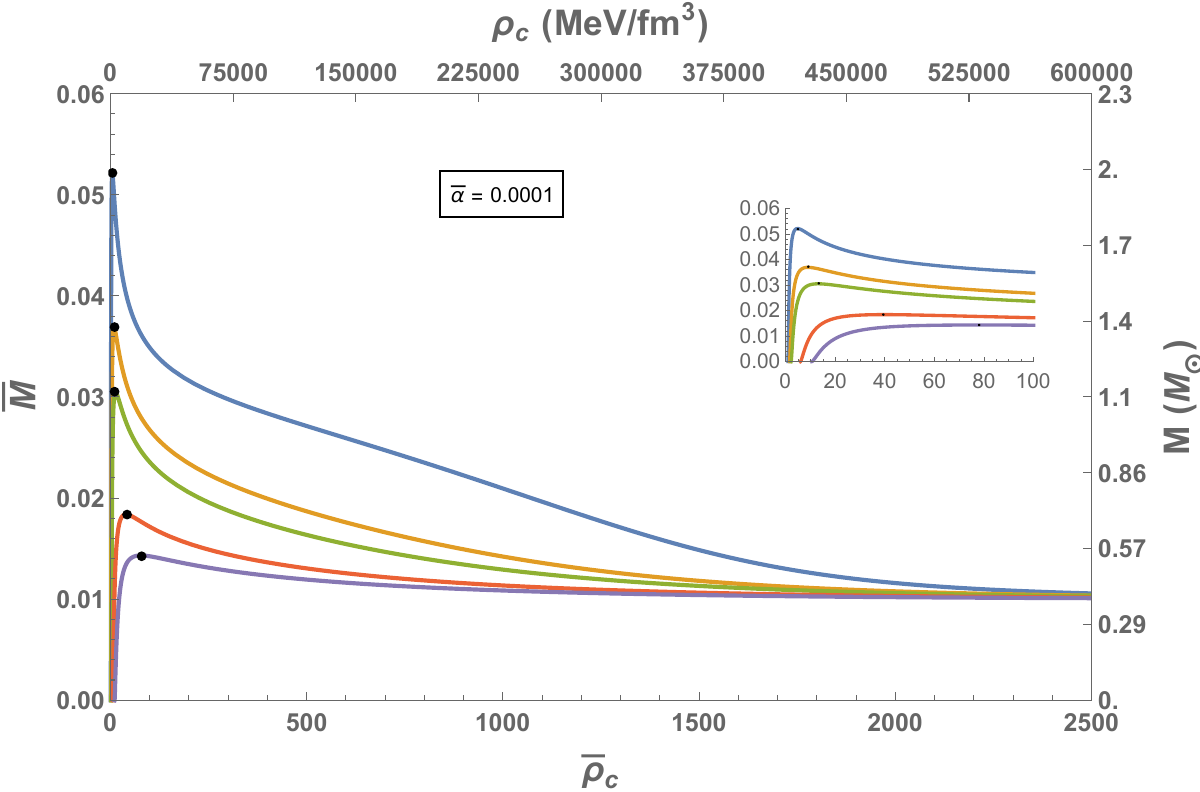}
        }

        \subfloat[\label{fig:tov apt001 neglam}]{
        \includegraphics[width=7.6cm]{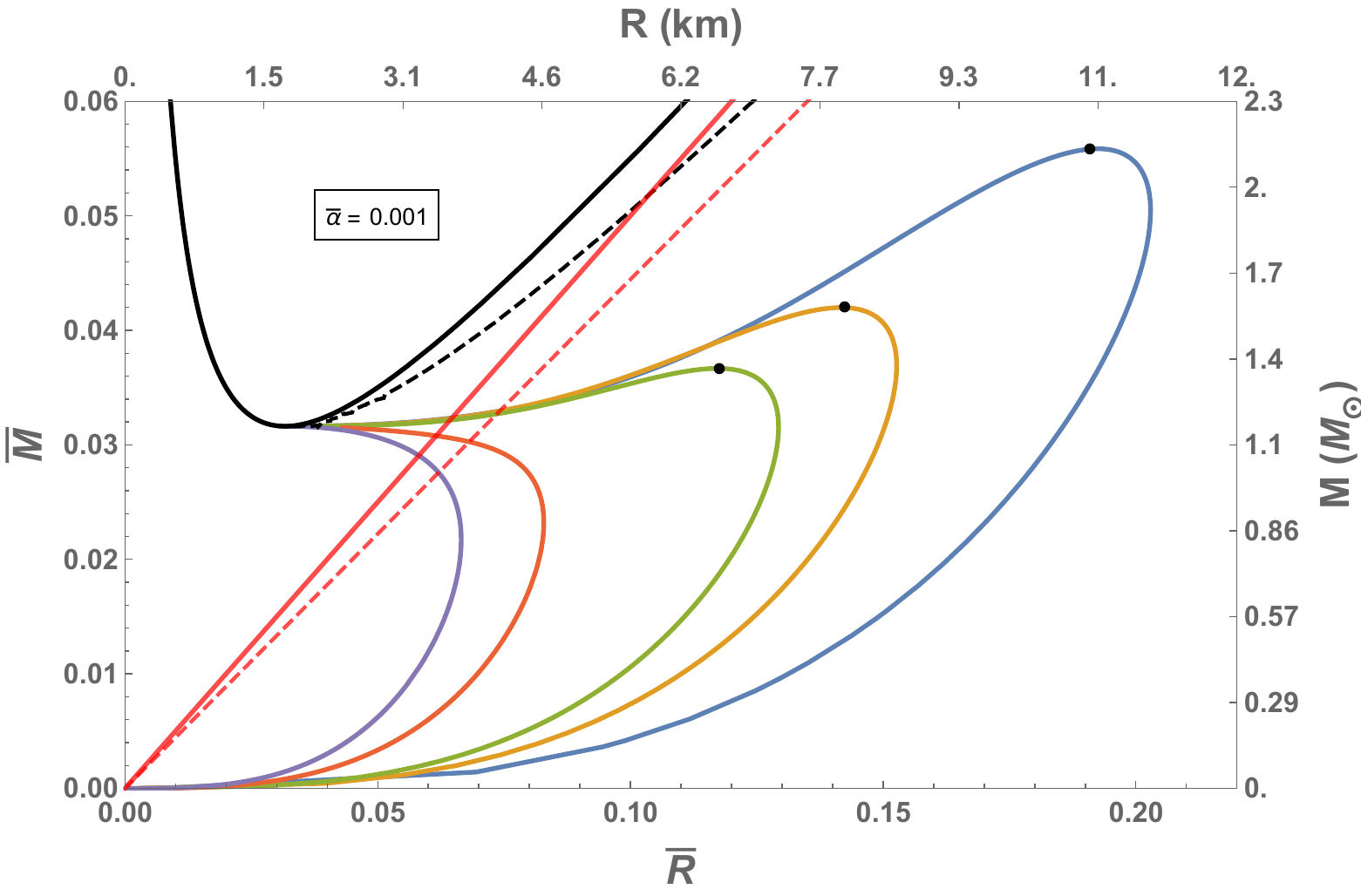}
        }\hfill
        \subfloat[\label{fig:mpc apt001 neglam}]{
        \includegraphics[width=7.6cm]{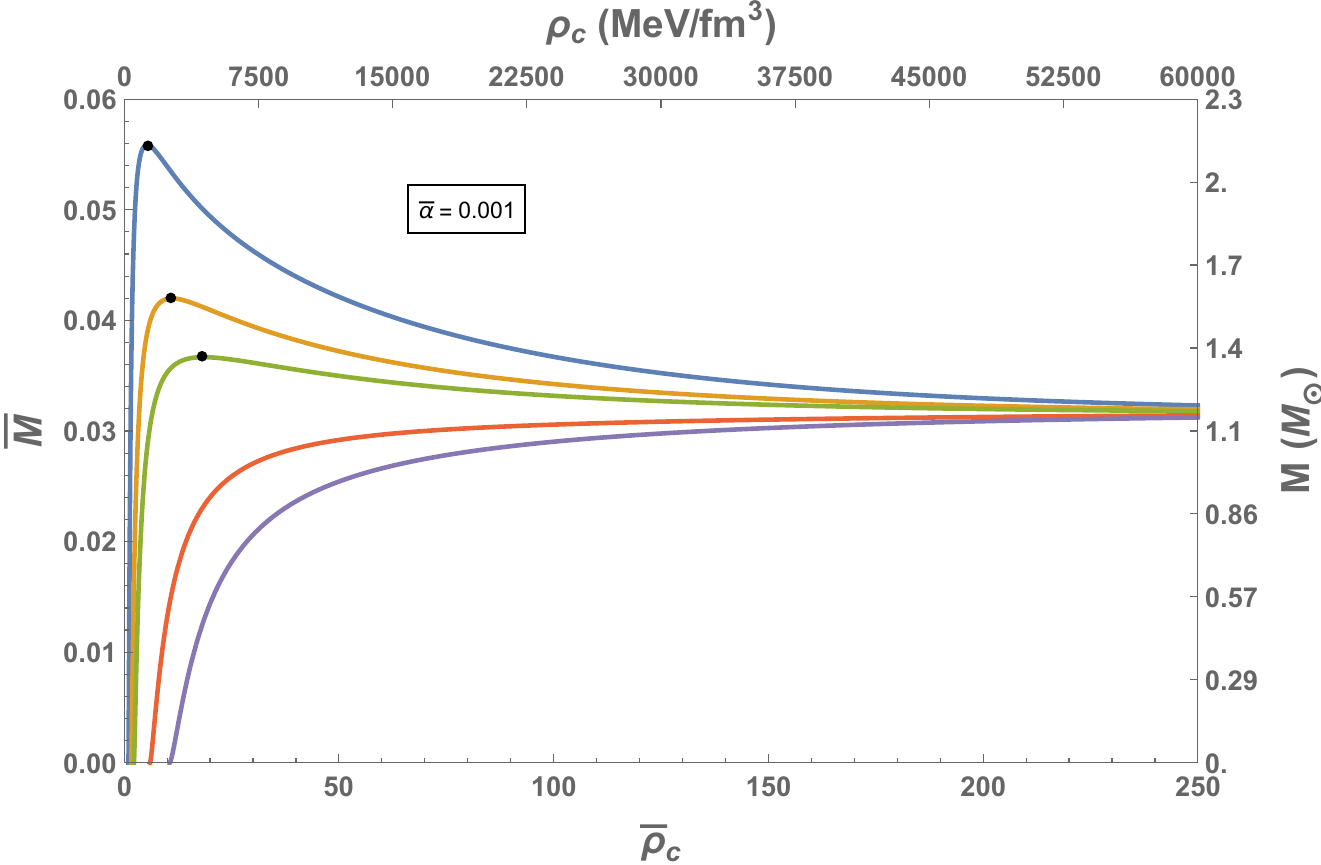}
        }

	\caption[]{Mass-radius and mass-central density relations in the case where $\lambda < 0$. Each $\bar{\alpha}$ corresponds to a different set of $\bar{\lambda}$ for which the solutions are plotted, due to the extreme sensitivity of the critical pressure with respect to negative $\lambda$. In all cases the ratio $\frac{\bar{\lambda}}{\bar{\lambda}_\mathrm{crit}}$ takes on the same set of values, namely 0, 1/25, 1/10, 1/2, 1 respectively from the blue to purple curves. The effect of increasing $\bar{\lambda}$ (decreasing $\lambda$) is to suppress the mass and radius, a logical extension of the $\lambda>0$ results. \label{fig:neglamsolutions6}}
\end{figure*}

\begin{figure*}
        \subfloat[\label{fig:tov apt01 neglam}]{
        \includegraphics[width=7.6cm]{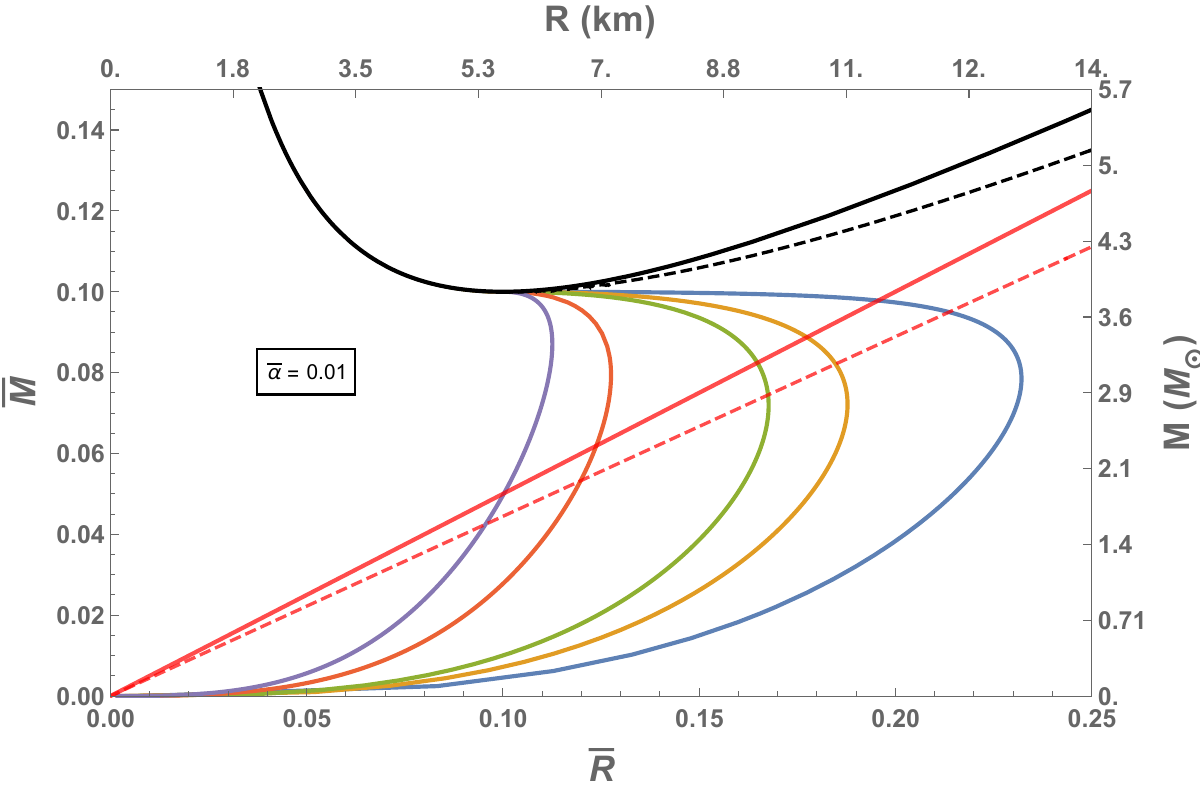}
        }\hfill
        \subfloat[\label{fig:mpc apt01 neglam}]{
        \includegraphics[width=7.6cm]{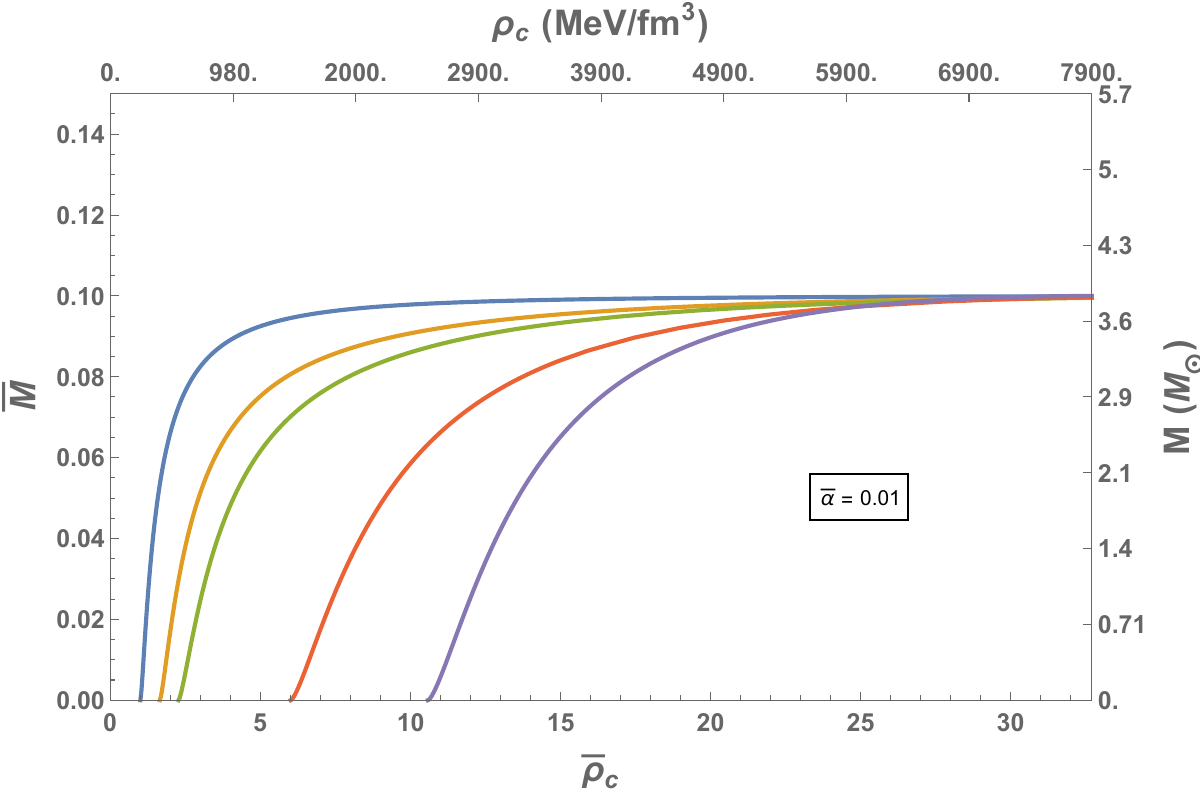}
        }

        \subfloat[\label{fig:tov apt1 neglam}]{
        \includegraphics[width=7.6cm]{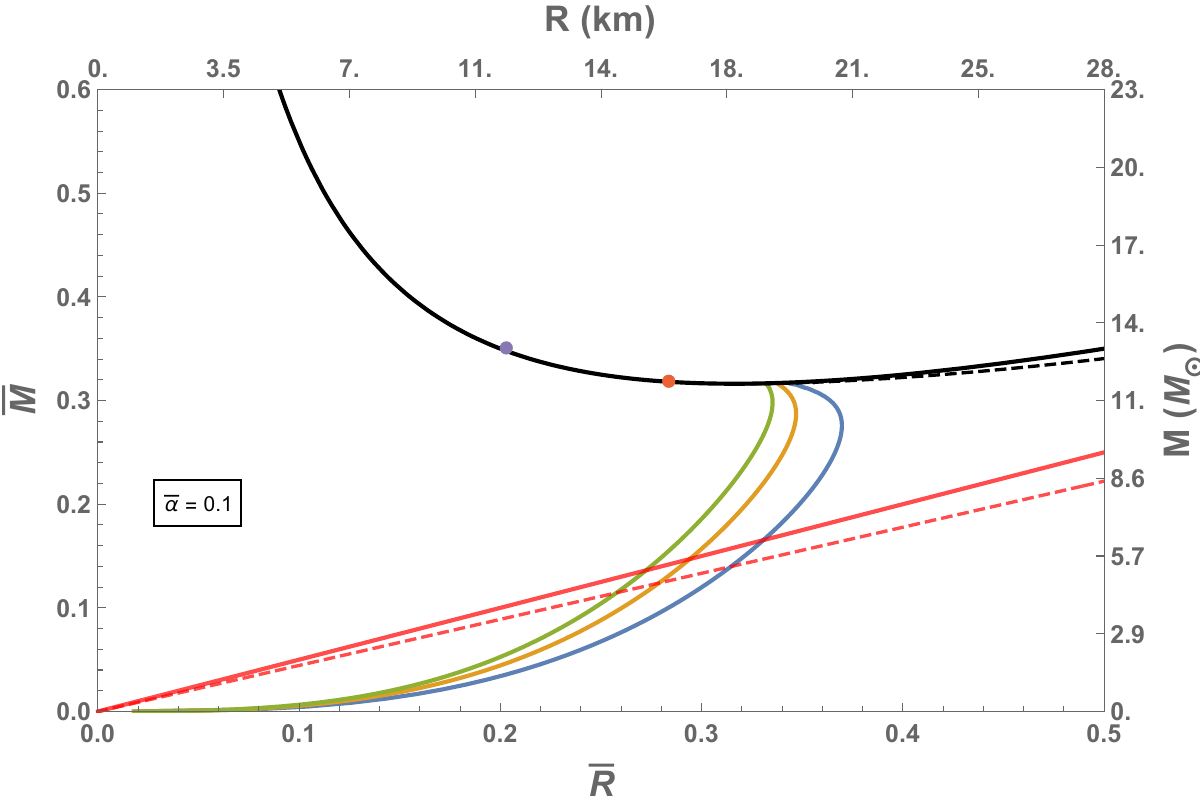}
        }\hfill
        \subfloat[\label{fig:mpc apt1 neglam}]{
        \includegraphics[width=7.6cm]{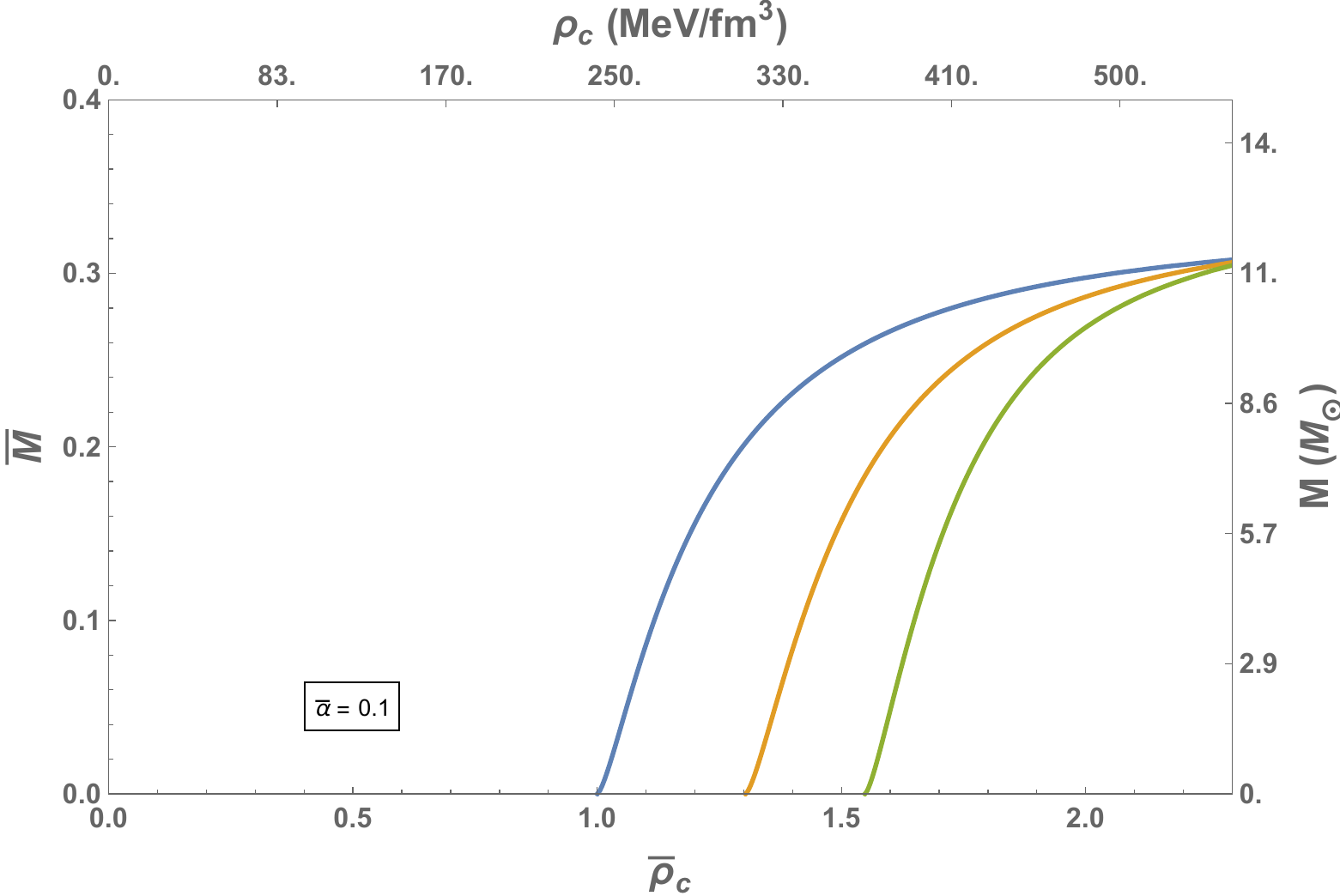}
        }

        \subfloat[\label{fig:tov apt25 neglam}]{
        \includegraphics[width=7.6cm]{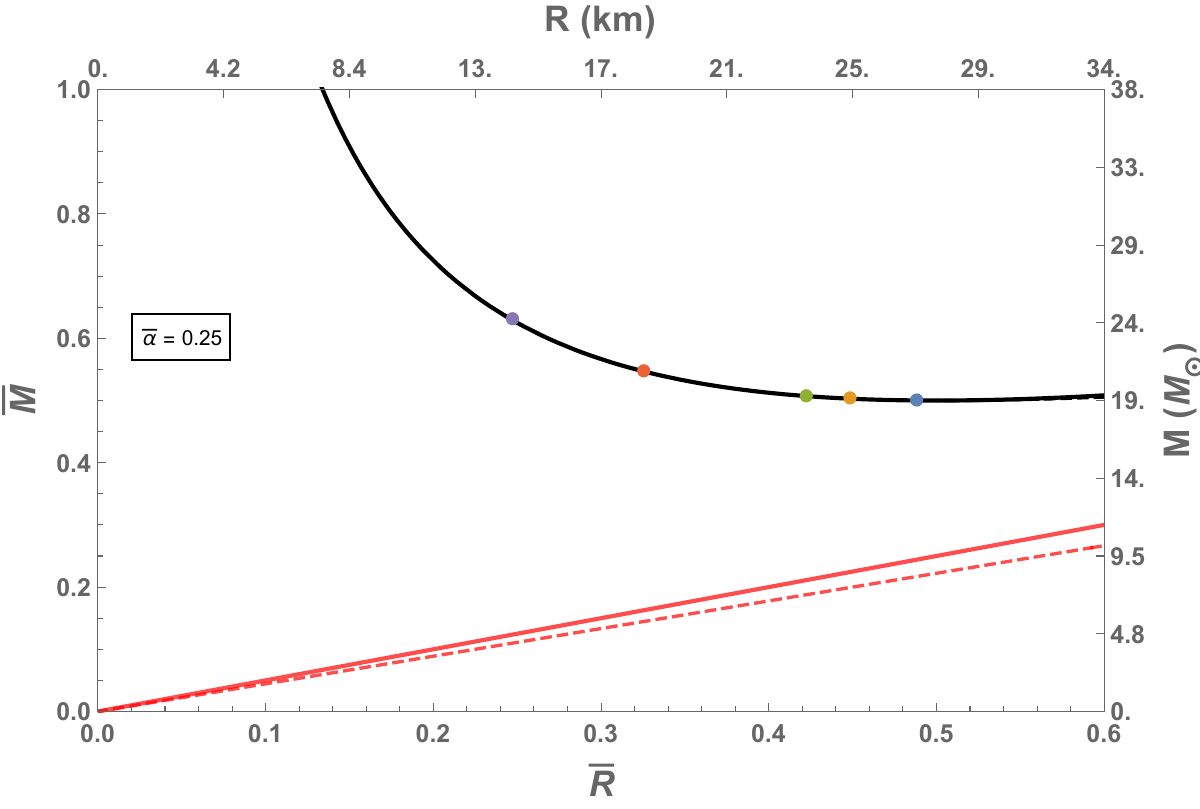}
        }\hfill
        \subfloat[\label{fig:mpc apt25 neglam}]{
        \includegraphics[width=7.6cm]{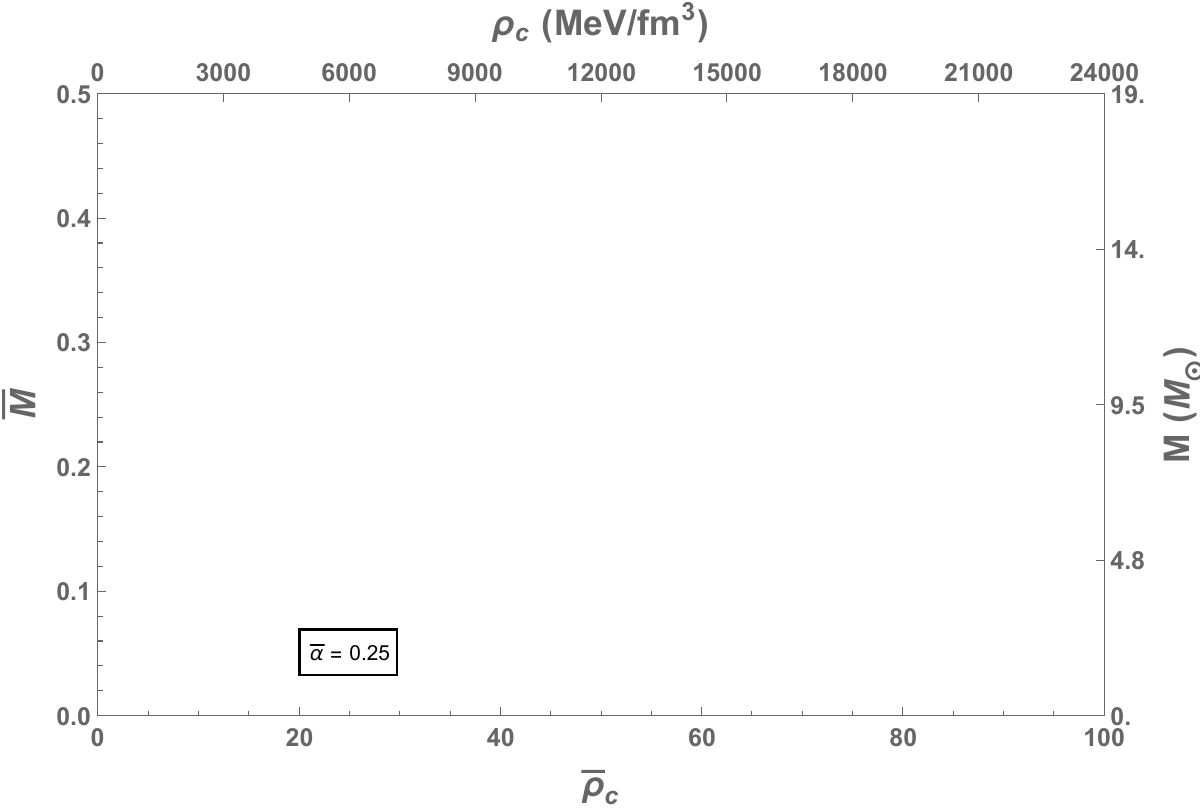}
        }

	\caption[]{Mass-radius and mass-central density relations in the case where $\lambda < 0$. Each $\bar{\alpha}$ corresponds to a different set of $\bar{\lambda}$ for which the solutions are plotted, due to the extreme sensitivity of the critical pressure with respect to negative $\lambda$. In all cases the ratio $\frac{\bar{\lambda}}{\bar{\lambda}_\mathrm{crit}}$ takes on the same set of values, namely 0, 1/25, 1/10, 1/2, 1 respectively from the blue to purple curves. The effect of increasing $\bar{\lambda}$ (decreasing $\lambda$) is to suppress the mass and radius, a logical extension of the $\lambda>0$ results. \label{fig:neglamsolutions7}}
\end{figure*}

Here we summarize the results for   $\lambda < 0$, which is equivalent to the condition
\begin{equation}\label{eq:neglamcondition}
    \xi_{2 a} \Delta^2<\xi_{2 b} m_s^2.
\end{equation}
In this case (and so throughout this section) a larger $\bar{\lambda}$ corresponds to a negative $\lambda$ that is larger in magnitude (more negative). As is shown in detail in appendix \ref{sec:appendixcritpres}, the critical pressure now has a divergence at $\bar{\alpha} = \frac{9 \pi}{32 \bar{\lambda}}$ in this regime. This defines a critical relationship between the 4D Gauss-Bonnet coupling and the strong interaction parameter at which the central pressure must approach infinity for a valid quark star solution. This also means that the allowed range of interaction strengths inside the star vary based on the choice of coupling to the higher curvature term. 

We illustrate our results in
Figures~\ref{fig:neglamsolutions6} and~\ref{fig:neglamsolutions7}. 
In order to keep the results manageable and easy to interpret, we choose to never use $\bar{\alpha}-\bar{\lambda}$ combinations that result in a critical central pressure larger than $\bar{p}_\mathrm{crit}=10$ (and define $\bar{\lambda}_\mathrm{crit}$ to be that which corresponds with this solution set, since $\bar{\lambda}_\mathrm{max}$ has a divergent critical pressure - see Figure \ref{fig:critical pressure negative lambda}). By doing this, we can more easily compare a range of $\bar{\alpha}$ solutions with the same choices of $\frac{\bar{\lambda}}{\bar{\lambda}_\mathrm{crit}}$, and view them on axes of comparable magnitudes.

As expected we see that decreasing $\lambda$ (and hence increasing $\bar{\lambda}$) causes the mass and radius of a given solution to trend downward, whereas increasing $\bar{\alpha}$ yields the same upward trend as before. In other words, in the negative lambda regime strong interaction effects tend to somewhat counteract the effects of non-zero coupling to the Gauss-Bonnet theory. We also notice that the effects of the critical pressure are first manifest on the curves with a larger negative magnitude, and thus the end of the solution space occurs when $\alpha$ is large enough to trigger criticality for the case of vanishing interaction effects ($\bar{\lambda} = 0$). As in the previous section this will occur when $\bar{\alpha} = \frac{3}{4 \pi}$, and thus a stronger coupling to 4DEGB theory will result in empty plots as seen in Figures \ref{fig:tov apt25 neglam} and \ref{fig:mpc apt25 neglam}.

Finally, we note from Figure~\ref{fig:neglamsolutions6} that there are again regions where solutions exist that violate the GR Buchdahl bounds and Schwarzschild radius, even for very small values of $\bar{\alpha}$. The range of solutions gets large as $\bar{\alpha}$ increases, as shown in Figure~\ref{fig:neglamsolutions7}.

\section{Stability Analysis}\label{sec:stability}

\begin{figure*}
\subfloat[Zoomed in.\label{fig:causality}]{
        \includegraphics[width=7.6cm]{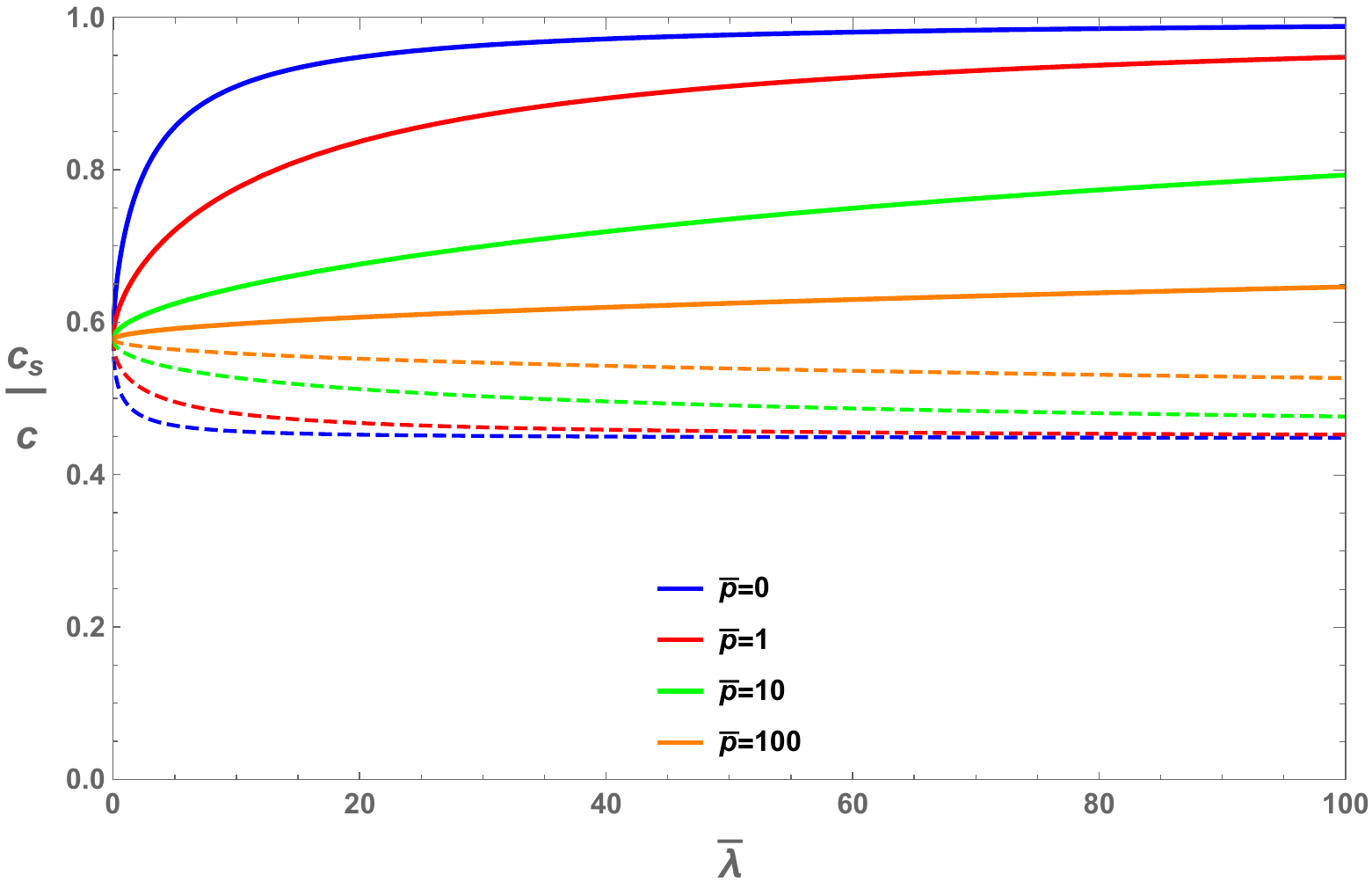}
        }\hfill
        \subfloat[Zoomed out.\label{fig:causality stronginteraction}]{
        \includegraphics[width=7.6cm]{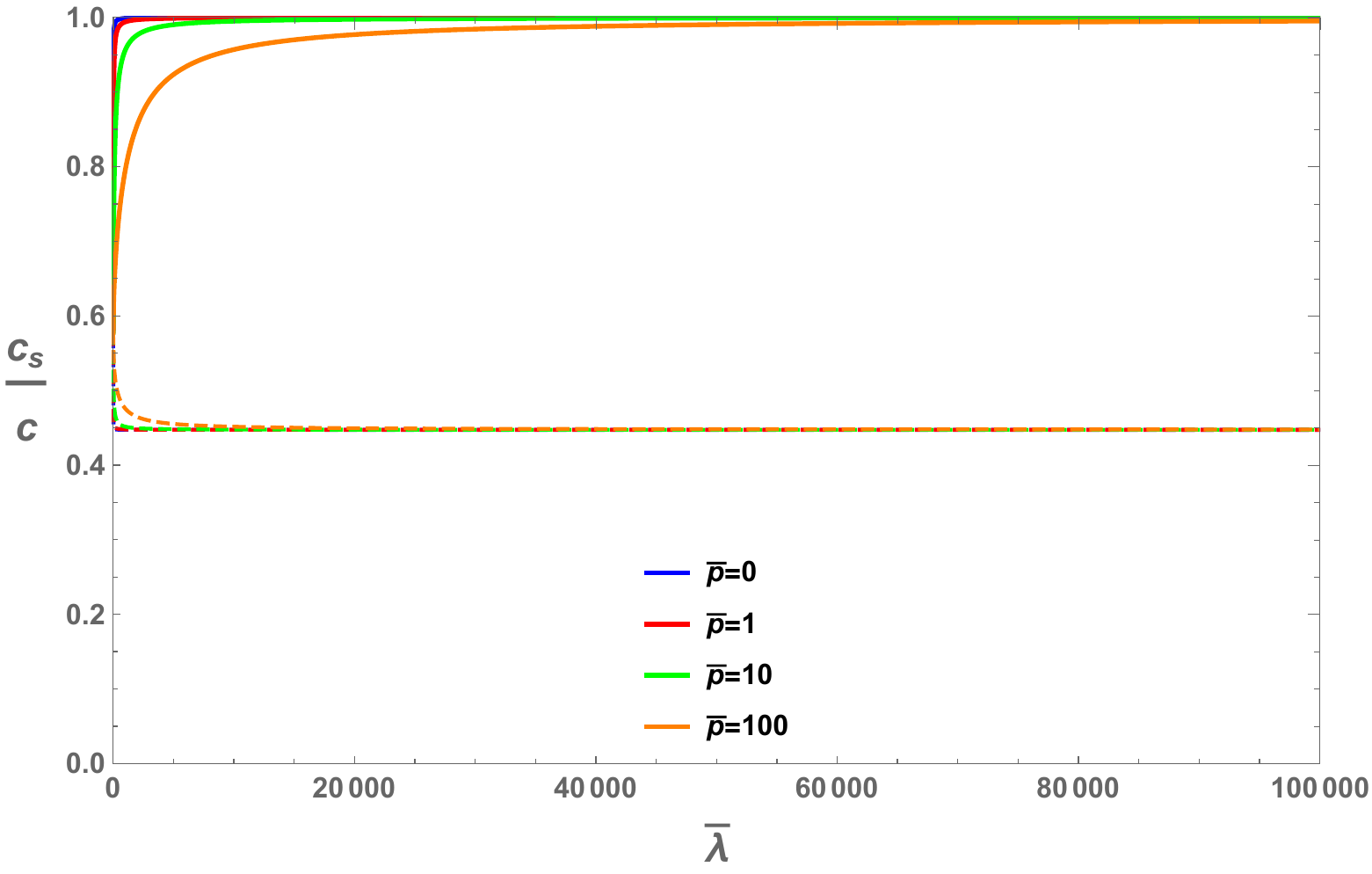}
        }
        
	\caption[]{Sound speed inside the unified interacting quark star as a function of interaction strength $\bar{\lambda}$ when pressure is fixed (note that this quantity is independent of the coupling strength $\alpha$). The solid lines correspond to $\lambda>0$ and the dashed lines $\lambda<0$. \label{fig:bothcausality}}
\end{figure*}

We now consider the stability of the 4DEGB unified interacting quark stars. A necessary but insufficient condition for an uncharged compact star's stability in Einstein gravity is $d M/d \rho_c< 0$ \cite{glendenning1998_nonidentical,arbanil2016_radial,zhang_2021_stellar}, corresponding to the parts of the solution curves before maximum mass points are reached. In Einstein-Maxwell theory a net charge can offset stability from the maximum mass point in either direction \cite{zhang_2021_unified}. In a similar vein, when the 4DEGB theory coupling is non-zero, it is not obvious whether the coincidence of stability and maximum mass point will hold. Leaving a thorough analysis of the fundamental radial oscillation modes  for future work, we note that stable compact objects must obey the causality condition  that speed of sound $c_s$ never exceeds the speed of light $c$. Inserting the equation of state \eqref{eq:eos} directly into the definition for $c_s$, we find 
\begin{equation}
    c_s = \left(3\; -\; \frac{\textrm{sgn}({\lambda})4}{\sqrt{\frac{4 \lambda +4 \pi ^2 \bar{p}(\bar{r})+\pi ^2}{\lambda }}}\right)^{-1/2}\; .
\end{equation}
To attain the limit $c_s \to 1$ from below  implies that $\bar{p}(r) = -1/4$ (and $\lambda>0$), a  value that is never part of the quark star solution space. In Figure \ref{fig:bothcausality} we plot the sound speed as a function of the interaction strength $\lambda$ for various fixed values of the pressure. We see that for positive pressures the $c_s \to 1$ limit is asymptotically approached from below as $\lambda \to \infty$; the sound speed is never superluminal. If $\lambda<0$ instead, the analogous solutions curve downward and asymptotically approach $c_s = 1/\sqrt{5}$  as expected from \cite{zhang_2021_unified,zhang_2021_stellar}. 

Another indicator of stability is  the effective adiabatic index $\gamma$ of perturbations.  For adiabatic oscillations, this can be defined via the speed of sound through the quark-gluon plasma:
\begin{equation}
    \gamma_\mathrm{eff} \equiv\left(1+\frac{\rho}{P}\right)\left(\frac{d p}{d \rho}\right)_S
\end{equation}
where the subscript $S$ indicates that we consider the sound speed at constant specific entropy (recalling that $c_s = \sqrt{(\frac{\partial p}{\partial \rho})_S}$). One can consider this quantity to be a bridge ``between the relativistic structure of a spherical static object and the equation of state of the interior fluid" \cite{moustakidis2017_stability}. In principle a critical value for $\langle \gamma_\mathrm{eff} \rangle$ exists, below which configurations are unstable against radial perturbations. In standard general relativity this critical value can be written \cite{chandrasekhar1964,moustakidis2017_stability} $\gamma_{c r}=\frac{4}{3}+\frac{19}{42} \beta$, where $\beta=2 M / R = R_S/R$ is the compactness parameter. If $\beta \to 0$ the well-known classical Newtonian limit is recovered as expected ($\langle\gamma_\mathrm{eff}\rangle \geq \frac{4}{3}$).

\begin{figure*}
\subfloat[$\bar{\lambda}=0$\label{fig:gamma lam 0}]{
        \includegraphics[width=7.6cm]{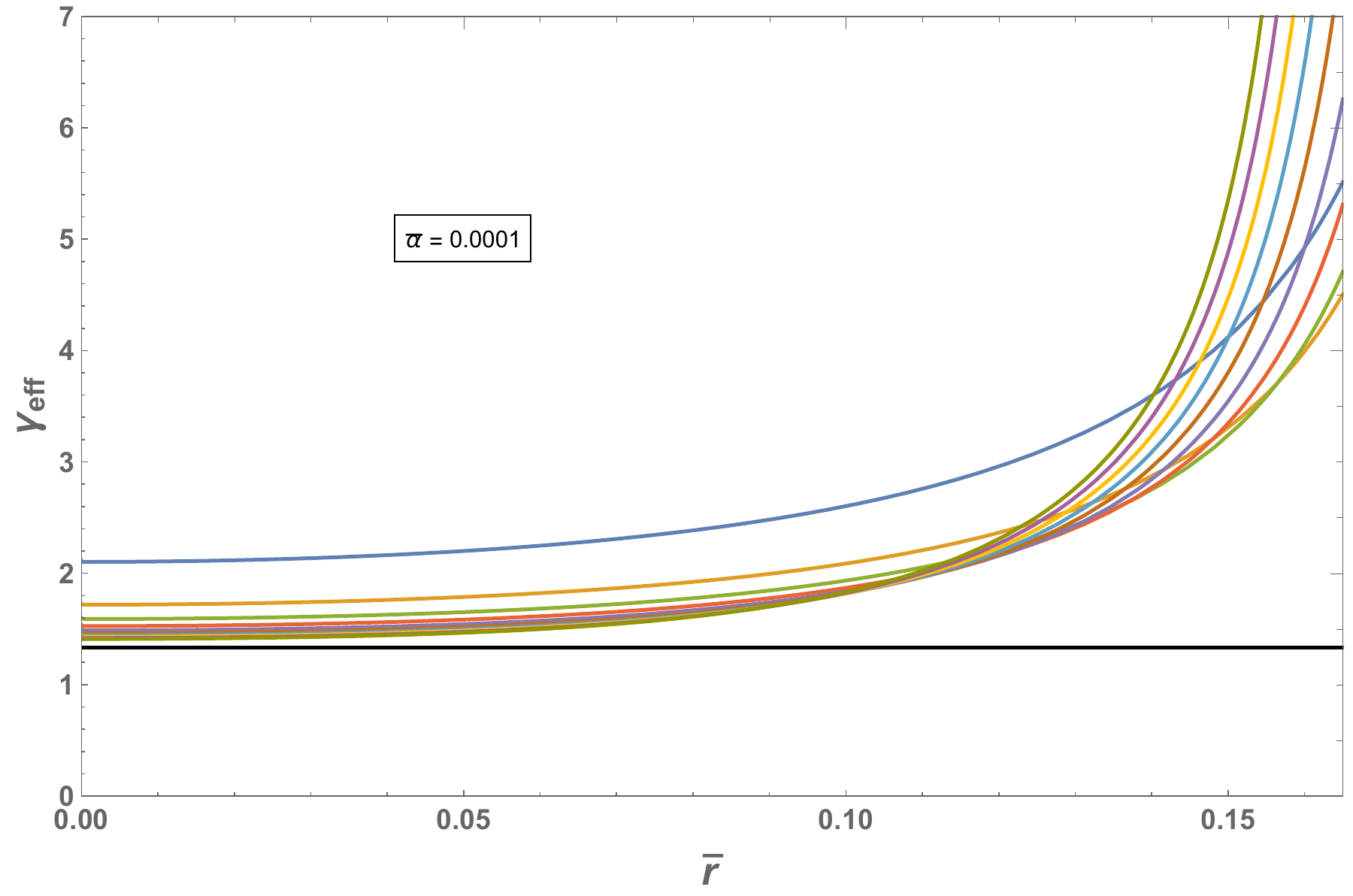}
        }\hfill
        \subfloat[$\bar{\lambda}=\infty$\label{fig:gamma lam inf}]{
        \includegraphics[width=7.6cm]{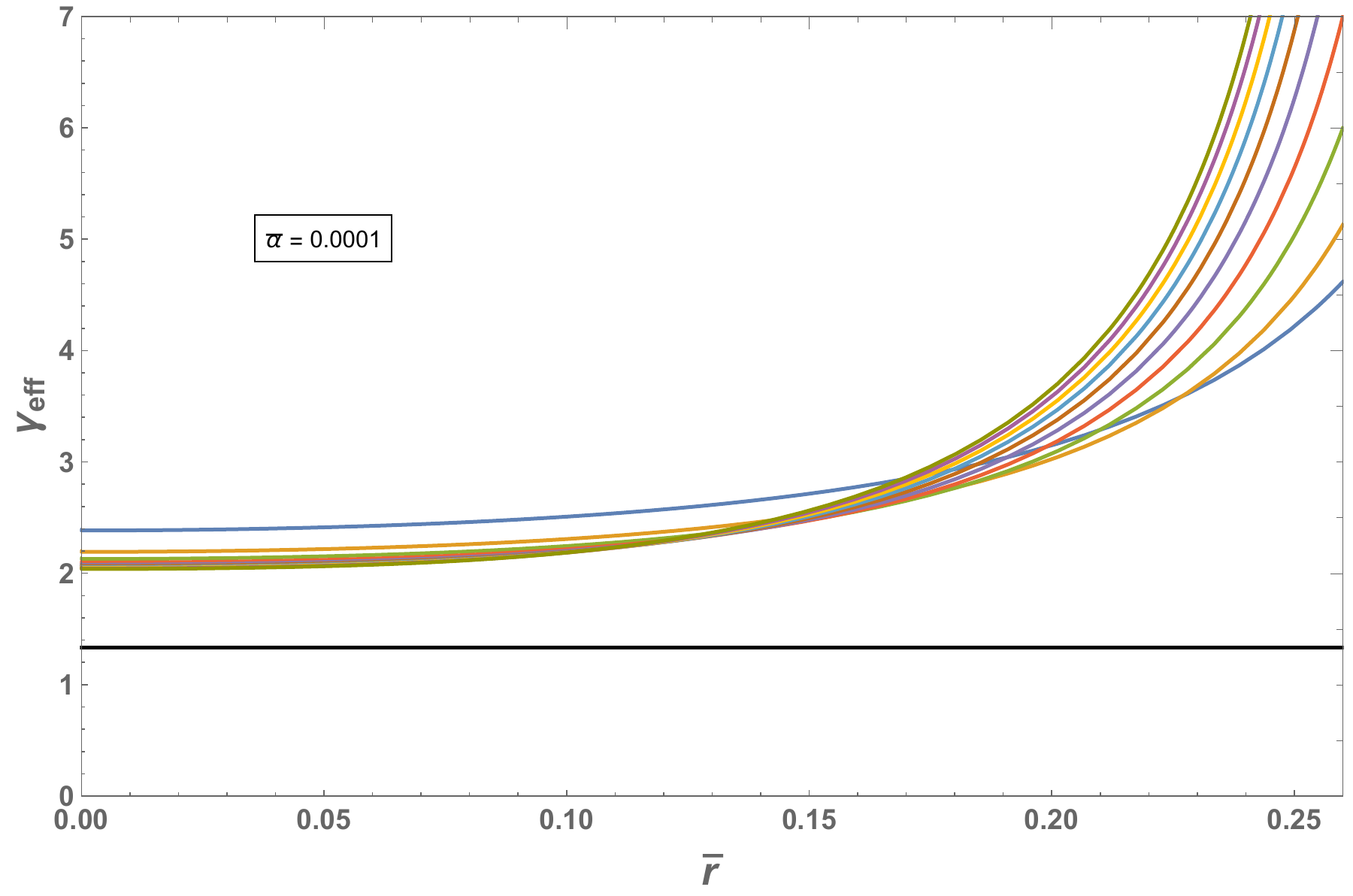}
        }
        
	\caption[]{Adiabatic index $\gamma_\mathrm{eff}$ plotted over the surface of the star when $\bar{\alpha} = 0.0001$ and $\lambda > 0$. Different coloured curves represent different central pressures (with the smallest central pressures having the largest $\gamma_\mathrm{eff}$ at the origin), and the black curve is the Newtonian limit 4/3. \label{fig:gammas}}
\end{figure*}

An equivalent bound has not yet been derived for the 4DEGB theory. Despite this, it is common practice to plot the adiabatic index of the star relative to the Newtonian critical value \cite{banerjee_2021_quark,banerjee_2021_strange,hansraj_2020_isotropic,singh_2022_anisotropic}. In general $\gamma$ will depend on $\lambda, \alpha, \rho_0$ and $r$. Given the size of our parameter space, plotting this would be cumbersome and not particularly illuminating (although an example case is shown in Figure \ref{fig:gammas}). We already know due to the form of the equations that $\lim_{r \to R}\gamma \to \infty$, and due to the monotonically decreasing nature of our pressure profiles, $\frac{\partial \gamma}{\partial r}>0$. With this, we are mostly interested in the lower limit of $\gamma_{r\to 0}$ (ie. where the curves in Figure \ref{fig:gammas} start from). This information is laid out in table \ref{tab:table} where, noting that $\lim_{p \to  \infty} \gamma_\mathrm{eff}$ is the lower limit on $\gamma_{r=0}$, 
we can see that in all cases $\gamma_{r=0} \geq \frac{4}{3}$, and also that $\gamma$ always diverges at the surface of the star. Since there is no well defined upper limit on interaction when $\lambda<0$, the first column of the third row cannot be further simplified. The results of this table imply that $\langle \gamma \rangle > \gamma_\mathrm{crit}^\mathrm{GR}$ in all cases. 

\begin{table}
\caption{Limits on the adiabatic index $\gamma$.}
\begin{tabular}{|l|l|l|l|}
\hline
                        & $\gamma_\mathrm{eff}$ & $\lim_{p \to \infty} \gamma_\mathrm{eff}$ 
                         & $\lim_{p \to  0} \gamma_\mathrm{eff}$ ($\gamma_{r \to R}$) \\ \hline
$\lambda = 0$           & $\frac{1}{3}(4 + \frac{1}{p(r)})$        & $\frac{4}{3}$                                                               & $\infty$                                           \\ \hline
$\lambda \to \infty$ & $2(1 + \frac{1}{4 p(r)})$       & $2$                                                              & $\infty$                                            \\ \hline
$\lambda < 0$           & $\frac{\sqrt{\frac{4 \lambda +4 \pi ^2 p(r)+\pi ^2}{\lambda }} \left(4 \lambda +2 \lambda  \sqrt{\frac{4 \lambda +4 \pi ^2 p(r)+\pi ^2}{\lambda }}+4 \pi ^2 p(r)+\pi ^2\right)}{\pi ^2 p(r) \left(3 \sqrt{\frac{4 \lambda +4 \pi ^2 p(r)+\pi ^2}{\lambda }}+4\right)}$      & $\frac{4}{3}$                                                             & $\infty$                                           \\ \hline
\end{tabular}
\label{tab:table}
\end{table}

\section{Summary}\label{sec:summary}

In this paper we have investigated the stellar structure of strongly interacting quark stars in the regularized 4D Einstein Gauss-Bonnet theory of gravity for different combinations of the 4DEGB coupling constant $\alpha$ and the unified strong interaction parameter $\lambda$ in our interacting quark matter equation of state.  In accord with the lack of a mass gap in the 4DEGB theory \cite{charmousis2022}, we find that -- for both signs of the coupling $\lambda$ -- even for small $\alpha$ the quark star solutions asymptotically approach the 4DEGB black hole horizon radius, and thus have solutions with smaller radii than the GR Buchdahl/Schwarzschild limits. It is worth noting that in the $\lambda<0$ case a much larger central pressure is required to approach this limit as compared to the analogous $\lambda>0$ case. In general, larger $\alpha$ and $\lambda$ tend to increase the mass-radius profile of quark stars, and a large negative $\lambda$ has the opposite effect, pushing back against the contributions of non-zero coupling to the higher curvature gravity terms. These findings are generally consistent with what was found in the regime of weak coupling to the 4DEGB theory   \cite{banerjee_2021_quark}. 

We have found  many additional features in the unexplored regions of parameter space, the most striking of which is that 4DEGB quark stars can exist not only below the GR Buchdahl bound, but also smaller than the $2M$ Schwarzschild radius. These objects obey the criterion that their sound speed is subluminal and have an average effective adiabatic index above what is required in GR. An interesting avenue for future work would be to study the stability properties of these objects, and of compact 4DEGB stars in general. 

We have also analytically identified a critical central pressure (which is a function of $\alpha$/$\lambda$ only), below which quark star solutions do not exist. This critical pressure cannot be physically realized; rather it corresponds to quark stars whose radii are either equal to or smaller than the corresponding black hole of the same mass. In principle, the $M-R$ pairs associated with this critical pressure can be used to constrain the constants of the theory (for instance, if a particular $\alpha$/$\lambda$ pair does not allow for $M-R$ pairs found observationally).

In addition to a detailed study of the stability of compact stars in 4DEGB gravity, our work should be extended to consider the effects of charge on stellar structure, since the confining nature of the strong interaction makes quark stars one of the few remaining candidates for stellar objects which are not electrically neutral \cite{zhang_2021_stellar,pretel_2022,gammon_2022_beyond}. Moreover, real astrophysical objects generally also have a net angular momentum, which can alter the stellar structure. Our calculations could be extended to a slowly rotating metric ansatz \cite{adair2020,gammon_2022} to see how the introduction of angular momentum changes the mass-radius relations of our interacting 4DEGB quark star. 

\section*{Acknowledgements}
This work was supported in part by the Natural Sciences and Engineering Research Council of Canada.  We are grateful to Sharon Morsink and Chen Zhang for helpful correspondence.




\bibliographystyle{unsrt}
\cleardoublepage 
\phantomsection  
\renewcommand*{\refname}{References}

\addcontentsline{toc}{chapter}{\textbf{References}}

\bibliography{refs}

\nocite{}


\appendix
\section{Critical Central Pressure}\label{sec:appendixcritpres}

\begin{figure*}[h]
        \subfloat[A ``typical" monotonically decreasing pressure profile for a quark star with a well-defined surface at $\bar{p}(\bar{r}) = 0$. \label{fig:noncritical}]{
        \includegraphics[width=7.6cm]{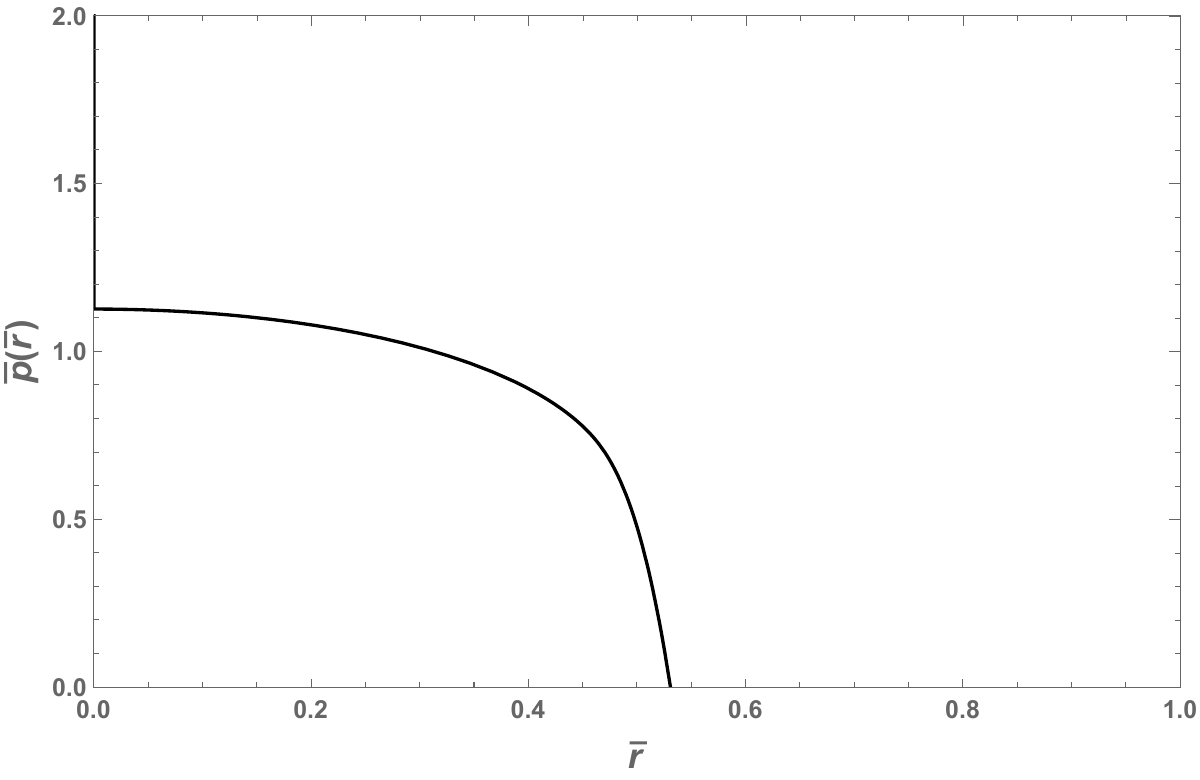}
        }\hfill
        \subfloat[$p_0 = p_c$ for our test case when $\lambda>0$. As expected we find a constant, flat line.\label{fig:crit}]{
        \includegraphics[width=7.6cm]{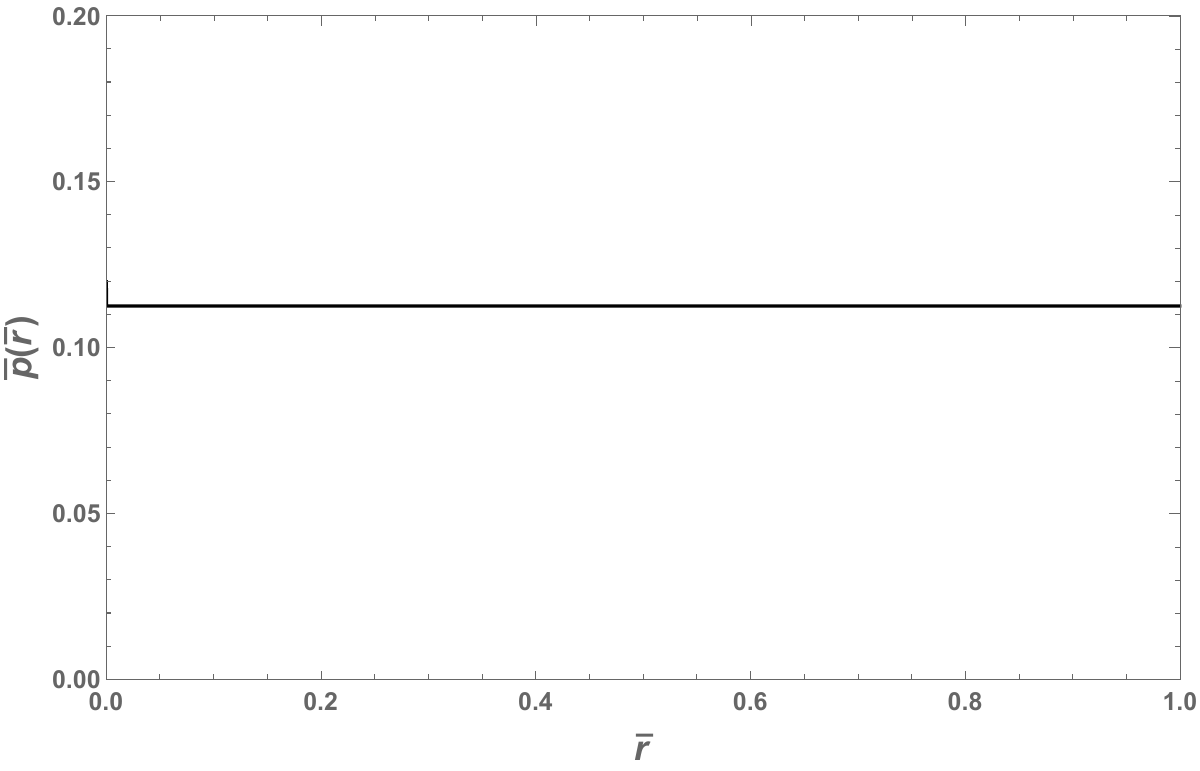}
        }
\caption{Critical central pressure analysis for $\lambda>0$.}
\end{figure*}

The quark star solution space is filled out numerically by scanning through test values for the star's central pressure $p_\mathrm{c}$ and finding the radius at which pressure vanishes to define the stellar surface. Typical compact star solutions have pressure profiles reminiscent of Figure \ref{fig:noncritical} where pressure decreases monotonically starting from $r = 0$. For particular $\alpha - \lambda$ combinations, no quark star solutions exist below some critical central pressure $p_\mathrm{crit}$. For $p_c$ below this critical value we find that the pressure blows up toward infinity and thus has no roots. In theory there should exist some $p_c = p_\mathrm{crit}$ whose pressure plot is constant, a straight horizontal line.

\begin{figure}
    \centering
    \includegraphics[width=0.8\textwidth]{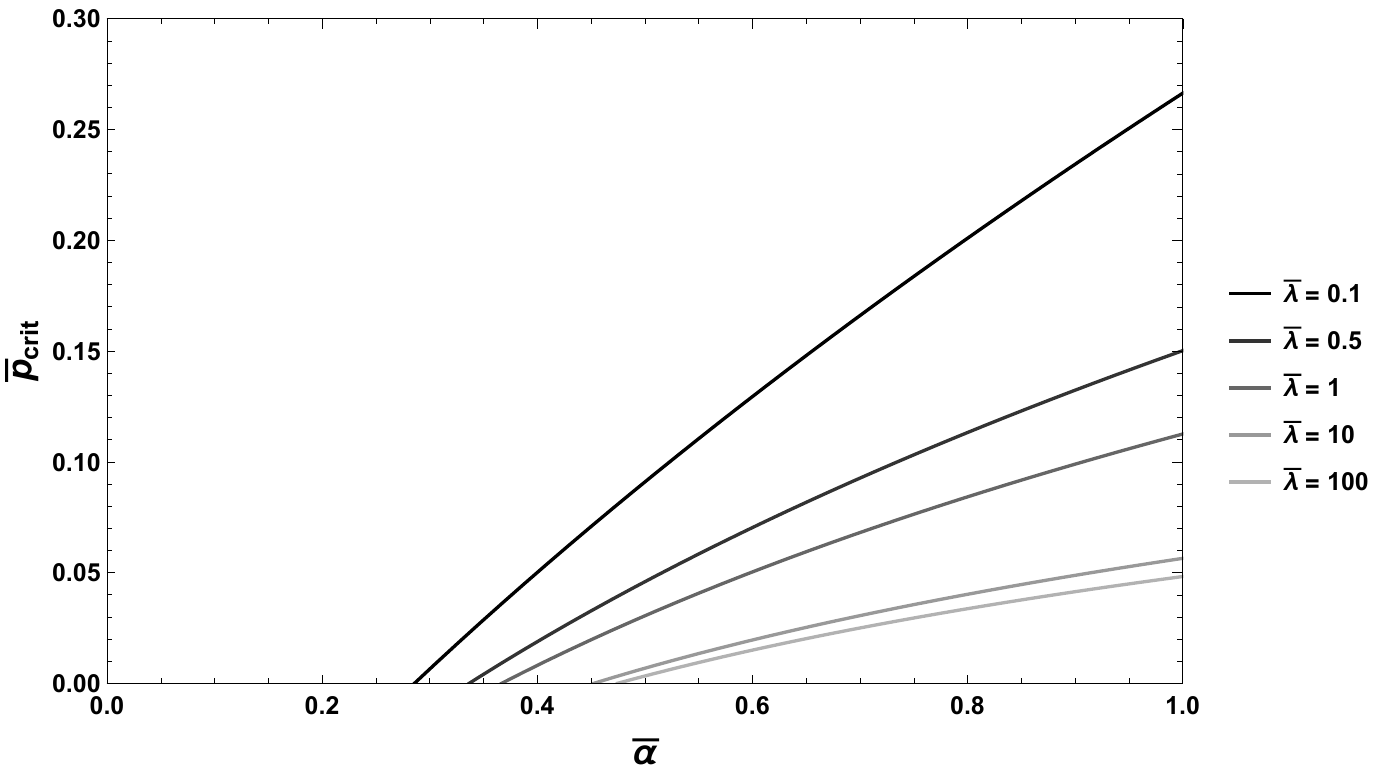}
    \caption{Critical pressure as a function of $\bar{\alpha}$ when $\lambda>0$, plotted for $\bar{\lambda} = 0.1, 0.5, 1, 10, 100$ (with the darkest lines corresponding to the smallest $\bar{\lambda}$). \label{fig:critical pressure positive lambda}}
\end{figure}

To derive this critical pressure, we  replace $\bar{p}(r)$ with the constant $\bar{p}_\mathrm{crit}$ in the EOS \eqref{eq:eosunitless}. The  
 differential equation \eqref{mdiff} for $m(r)$ becomes
 \begin{equation}\label{eq:mprimepcrit}
\bar{m}'(\bar{r}) = 4 \pi  \bar{r}^2 \left(-\frac{4 \bar{\lambda}  \left(\sqrt{\frac{\pi ^2 \left(\bar{p}_\mathrm{crit}+\frac{1}{4}\right)}{\lambda }+1}-1\right)}{\pi ^2}+3 \bar{p}_\mathrm{crit}+1\right)
\end{equation}
which is straightforwardly integrated, yielding
\begin{equation}\label{eq:msol}
\bar{m}(\bar{r}) = \frac{4 \bar{r}^3 \left(\pi ^2 (3 \bar{p}_\mathrm{crit}+1)-2 \bar{\lambda}  \left(\sqrt{\frac{4 \bar{\lambda} +4 \pi ^2 \bar{p}_\mathrm{crit}+\pi ^2}{\bar{\lambda} }}-2\right)\right)}{3 \pi } + c_1.
\end{equation}
where the boundary condition $\bar{m}(0) = 0$ implies $c_1 = 0$. Upon inserting this into the differential equation 
\eqref{pdiff} for  $\bar{p}(\bar{r})$
(and setting $\bar{p}'(\bar{r})=0$), we arrive at a condition corresponding to $\bar{p}(\bar{r}) = \bar{p}_c = \bar{p}_\mathrm{crit}$:
\begin{equation}
\begin{aligned}
    \bar{p}_\mathrm{crit} = \frac{1}{2 \pi  (9 \pi -32 \bar{\alpha}  \bar{\lambda} )^2} &\Bigg[256 \bar{\alpha}  (2 \pi  \bar{\alpha} +3) \bar{\lambda} ^2+8 \pi  (2 \pi  \bar{\alpha}  (8 \pi  \bar{\alpha} +21)-9) \bar{\lambda} +9 \pi ^3 (4 \pi  \bar{\alpha} -3)\\
    &-\textrm{sgn}(\lambda) 6 \sqrt{\bar{\lambda}  (32 \bar{\alpha}  \bar{\lambda} +\pi  (8 \pi  \bar{\alpha} -3))^2 \left(16 (2 \pi  \bar{\alpha} +1) \bar{\lambda} +\pi ^2 (8 \pi  \bar{\alpha} +3)\right)}\Bigg] .
    \end{aligned}  
\end{equation}
For  fixed  $\lambda > 0$, the critical pressure takes the form shown in Figure \ref{fig:critical pressure positive lambda}. If $\lambda =0$ then $\bar{p}_{crit} = \frac{4\pi\bar{\alpha}-3}{18}$, so the limiting intersection point is $\bar{\alpha}=\frac{3}{4\pi} = 0.238$.

For a given $\alpha - \lambda$ pair
only central pressures above $p_{crit}$ give valid quark star solutions. Since pressure must always be non-negative, a negative critical pressure 
implies that all choices of positive central pressure yield valid quark star solutions, and so to find solutions we can start scanning from an arbitrarily small positive pressure. However once $\alpha$ is sufficiently large, there is a threshold value of the central pressure below which there are no solutions.  The solutions in these cases when $p>p_\mathrm{crit}$ are found to \textit{start} at the black hole horizon and thus are never physically relevant. More simply, $\bar{\alpha}/\bar{\lambda}$ combinations for which $p_\mathrm{crit}>0$ have no physical quark star solutions, and thus in a sense this condition marks the physically relevant boundary of the parameter space. The requirement of a larger $p_c$ to avoid pressure divergence is somewhat counter-intuitive, although one can see from equation \eqref{eq:mprimepcrit} that a larger critical pressure corresponds to a larger mass which is more gravitationally attractive.

\begin{figure*}[h]
    \subfloat[$p_0 = p_\mathrm{crit} - \delta$.\label{fig:belowcrit}]{\includegraphics[width=7.6cm]{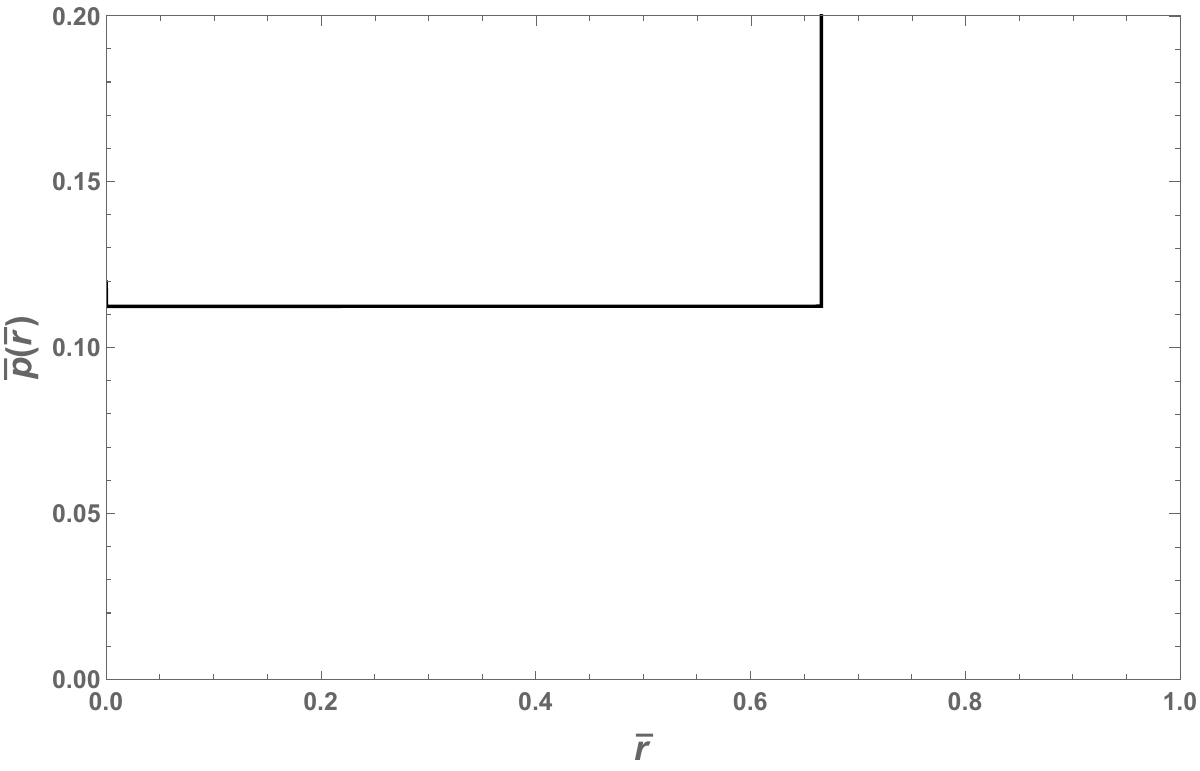}}\hfill
    \subfloat[$p_0 = p_\mathrm{crit} + \delta$.\label{fig:abovecrit}]{\includegraphics[width=7.6cm]{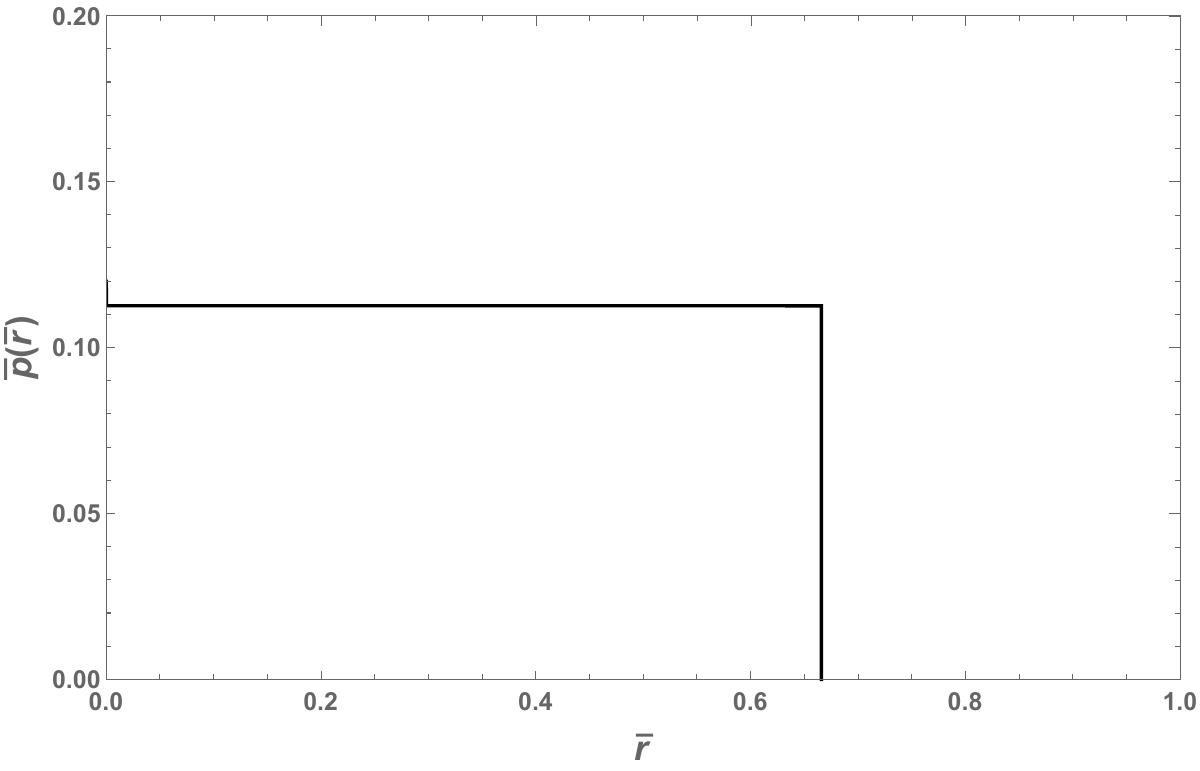}}
\caption{Either side of criticality for our test case when $\lambda>0$.}
\end{figure*}

To demonstrate this, for test values $\alpha = 1, \lambda = 1$ we find that $p_\mathrm{crit} = 0.11249510580691144$. We plot $p(r)$  for this trio of parameters in Figure \ref{fig:crit}, where as expected we see a flat line. If we instead we plot $p(r)$ for $p_0 = 0.112$ and $p_0 = 0.113$,
(figures \ref{fig:abovecrit} and  \ref{fig:belowcrit}),
we can see what happens on either side of criticality.




\begin{figure}
    \centering
    \includegraphics[width=0.8\textwidth]{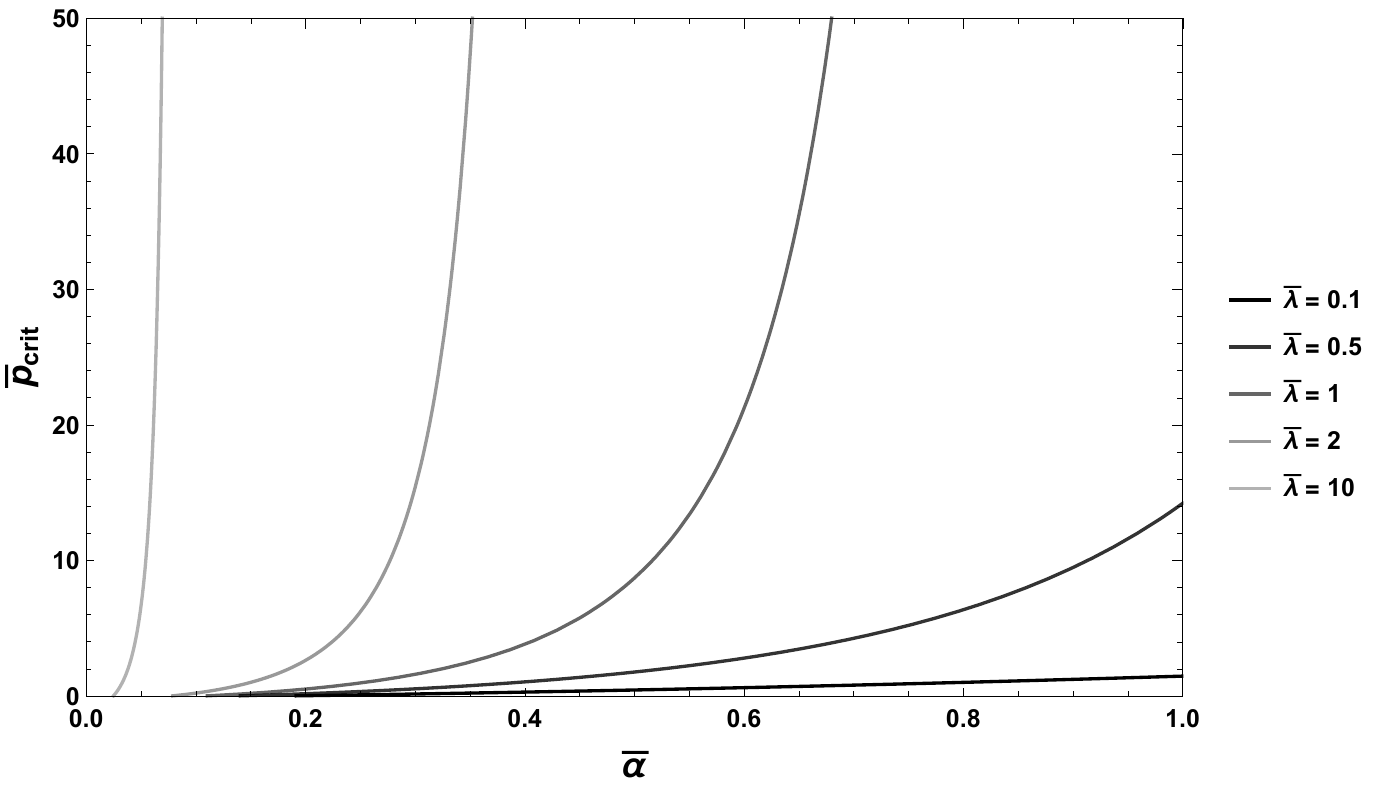}
    \caption{Critical pressure as a function of $\alpha$ when $\lambda<0$, plotted for $\bar{\lambda} = 0.1, 0.5, 1, 2, 10$ (with the darkest lines corresponding to the smallest $\bar{\lambda}$). Note that mathematically these functions past their discontinuity, though we find that this region does not produce quark star solutions and thus the corresponding curves have been removed (ie. the value of $\alpha$ at which $p_\mathrm{crit}$ diverges is the end of the solution space).}
    \label{fig:critical pressure negative lambda}
\end{figure}

On the other hand, when $\lambda<0$ 
the critical pressure solution has a divergence at $\bar{\alpha} = \frac{9 \pi}{32 \bar{\lambda}}$, which can be seen in Figure~\ref{fig:critical pressure negative lambda}. For values of $\bar{\alpha}$ smaller than the divergence point the critical pressure is defined analogously to the $\lambda>0$ case (namely only values above the curve are allowed). Past the divergence point there are no quark star solutions for any central pressures, and thus in the $\lambda<0$ regime we can say that $\bar{\alpha}_{\mathrm{max}} = \frac{9 \pi}{32 \bar{\lambda}}$.


\end{document}